\documentclass[numberedappendix]{emulateapj}
\usepackage{epsfig}
\usepackage{graphicx}
\usepackage{color}
\usepackage{enumerate}

\newcommand{\msun}{\,\rm M_\odot}

\newcommand{\be}{\begin{equation}}
\newcommand{\ee}{\end{equation}}
\newcommand{\ba}{\begin{eqnarray}}
\newcommand{\ea}{\end{eqnarray}}
\newcommand{\f}{\frac} 
\newcommand{\rvir}{r_{\rm vir}}
\newcommand{\rtwo}{r_{\rm 200}}
\newcommand{\mvir}{M_{\rm halo}}
\newcommand{\msub}{M_{\rm sub}}
\newcommand{\Vmax}{V_{\rm max}}
\newcommand{\rVmax}{r_{\rm Vmax}}
\newcommand{\rsub}{r_{\rm sub}}

\newcommand{\kms}{\,{\rm km\,s^{-1}}}

\newcommand{\MX}{M_\chi}
\newcommand{\mcut}{m_0}
\newcommand{\sigv}{\langle \sigma v \rangle}

\newcommand{\sig}{\mathcal{S}}

\lefthead{}
\righthead{The Dark Matter Annihilation Signal from Galactic Substructure: Predictions for GLAST}

\begin{document}
\submitted{}

\title{The Dark Matter Annihilation Signal from Galactic Substructure: \\Predictions for GLAST}

\author{Michael Kuhlen$^1$, J\"urg Diemand$^{2,3}$, Piero Madau$^{2}$}

\affil{$^1$School of Natural Sciences, Institute for Advanced Study,
Einstein Lane, Princeton, NJ 08540 \\$^2$Department of Astronomy \&
Astrophysics, University of California, Santa Cruz, CA
95064\\$^3$Hubble Fellow}

\begin{abstract}
  We present quantitative predictions for the detectability of
  individual Galactic dark matter subhalos in gamma-rays from dark
  matter pair annihilations in their centers. Our method is based on a
  hybrid approach, employing the highest resolution numerical
  simulations available (including the recently completed one billion
  particle Via Lactea II simulation) as well as analytical models for
  the extrapolation beyond the simulations' resolution limit. We
  include a self-consistent treatment of subhalo boost factors,
  motivated by our numerical results, and a realistic treatment of the
  expected backgrounds that individual subhalos must outshine. We show
  that for reasonable values of the dark matter particle physics
  parameters ($\MX \sim 50 - 500$ GeV and $\sigv \sim 10^{-26} -
  10^{-25}$ cm$^3$ s$^{-1}$) GLAST may very well discover a few, even
  up to several dozen, such subhalos, at 5 $\sigma$ significance, and
  some at more than 20 $\sigma$. We predict that the majority of
  luminous sources would be resolved with GLAST's expected angular
  resolution. For most observer locations the angular distribution of
  detectable subhalos is consistent with a uniform distribution across
  the sky. The brightest subhalos tend to be massive (median $\Vmax$
  of 24 $\kms$) and therefore likely hosts of dwarf galaxies, but many
  subhalos with $\Vmax$ as low as 5 $\kms$ are also visible.
  Typically detectable subhalos are $20 - 40$ kpc from the observer,
  and only a small fraction are closer than 10 kpc. The total number
  of observable subhalos has not yet converged in our simulations, and
  we estimate that we may be missing up to 3/4 of all detectable
  subhalos.
\end{abstract}

 
\section{Introduction}

Revealing the nature of dark matter is fundamental to cosmology and
particle physics. In the standard cosmological paradigm of structure
formation ($\Lambda$CDM), the universe is dominated by cold,
collisionless dark matter (CDM), and endowed with initial density
perturbations via quantum fluctuations during inflation. In this model
galaxies form hierarchically, with low-mass objects (``halos'')
collapsing earlier and merging to form larger and larger systems over
time. Small halos collapse at high redshift when the universe is very
dense, so their central densities are correspondingly high. When these
halos merge into larger hosts, their high densities allow them to
resist the strong tidal forces that act to destroy them. It is
therefore a clear, unique prediction of $\Lambda$CDM that galaxies are
embedded in massive, extended dark matter halos teeming with
self-bound substructure or ``subhalos''.

If the dark matter is in the form of a relic particle once in thermal
equilibrium in the early universe, like the neutralino in SUSY, then
substructures will be lit up by pair annihilations of these particles
into standard model particles, whose subsequent decay results in
gamma-ray emissions. The signal depends on unknown conjectured
particle physics properties and poorly known astrophysical
parameters. Particle physics uncertainties include the type of
particle (axion, neutralino, Kaluza-Klein particle, etc.), its mass,
and its pair annihilation cross-section. From the astrophysical side,
since the annihilation rate is proportional to density squared, the
predicted flux depends sensitively on the clumpiness of the mass
distribution.

The detection of annihilation radiation from DM clumps is surely one
of the most exciting discoveries that the upcoming \textit{Gamma-ray
  Large Area Space Telescope} (GLAST) could make. Currently scheduled
to be launched in May of 2008, GLAST is NASA's successor to the
Compton Gamma Ray Observatory, whose EGRET instrument conducted the
first all-sky gamma-ray survey and discovered 271 sources, 170 of
which remain unidentified \citep{Hartman1999}. GLAST will carry two
instruments, the GLAST Burst Monitor (GBM), designed to detect flashes
from gamma-ray bursts and solar flares, and the Large Area Telescope
(LAT), the main survey instrument. The LAT consists of several layers
of high-precision silicon tracking detectors for determining the
direction of an incident gamma-ray, a cesium-iodide calorimeter to
measure its total energy, and an anticoincidence detector for cosmic
ray rejection.

\begin{table*}[ht]
\begin{center}
\caption{Simulation Parameters}
\begin{tabular}{lcccccccccc}
\hline \hline
Name & $L_{\rm box}$ & $\epsilon$ & $z_i$ & $N_{\rm hires}$ & $M_{\rm hires}$ & $r_{200}$ & $M_{200}$ & $\Vmax$       & $\rVmax$ & $N_{\rm sub}$ \\
     & (Mpc)         & (pc)       &       &                 & ($\msun$)       & (kpc)     & ($\msun$) & (km s$^{-1}$) & (kpc)    &               \\
\hline
VL-I & 90.0 & 90.0 & 48.4 & $2.34 \times 10^8$ & $2.1 \times 10^4$ & 389 & $1.77 \times 10^{12}$ & 181 & 69 & 9,224 \\
VL-II & 40.0 & 40.0 & 104.3 & $1.09 \times 10^9$ & $4.1 \times 10^3$ & 402 & $1.93 \times 10^{12}$ & 201 & 60 & 53,653 \\
\hline
\end{tabular}
\tablecomments{Box size $L_{\rm box}$, (spline) softening length
  $\epsilon$, initial redshift $z_i$, number $N_{\rm hires}$ and mass
  $M_{\rm hires}$ of high resolution dark matter particles, host halo
  $r_{200}$, $M_{200}$, and $V_{\rm max}$, and number of subhalos
  within $r_{200}$ $N_{\rm sub}$ for the VL-I and VL-II
  simulations. Force softening lengths $\epsilon$ are constant in
  physical units back to $z=9$ and constant in comoving units
  before. }
\label{table:simulations}
\end{center}
\end{table*}

With a field of view of about 2.4 sr and a 90 minute orbital period,
the LAT will survey the entire sky daily. Its peak effective area
exceeds 8000 cm$^2$ beyond 1GeV and its angular resolution ranges from
$<3.5^\circ$ at 100 MeV down to $<9$ arcmin above 10 GeV. Compared to
EGRET, the LAT has a 4-5 times larger field of view, more than 5 times
larger peak effective area, 5-40 times higher angular resolution above
1 GeV, and $\sim 30$ times better point source
sensitivity. Additionally, the LAT detector is sensitive to gamma-rays
out to 300 GeV, ten times higher than EGRET's limit. Given these
improvements it is not unreasonable to hope that GLAST will discover
previously unknown sources, one of which may be annihilating DM in the
centers of Galactic subhalos.

What region in the sky is mostly likely to allow for a detection of
gamma-rays from DM annihilations is a hotly contested question. The
Galactic center (GC) is likely the largest nearby DM overdensity and a
number of studies have advocated it as the best place to look for a
signal \citep{Berezinsky1994, Bergstroem1998, Bergstroem2001,
  Ullio2002, Cesarini2004}. A major drawback for this scenario,
however, is the large astrophysical gamma-ray background near the GC.
Furthermore, the lack of a spectral break below $\sim 30$ TeV in the
recent H.E.S.S. \citep{Aharonian2006} and MAGIC \citep{Albert2006}
measurements of the very high energy gamma-ray emission from the GC
severely limits the contribution that annihilating DM particles with
masses less than $\mathcal{O}$(10 TeV) can make to this signal.
Instead, \citet{Stoehr2003} suggested to focus observations on an
annulus between 25$^\circ$ and 35$^\circ$ from the Galactic center,
since this would maximize the chances of detection by reducing the
background while simultaneously avoiding numerical uncertainties in
the central DM density profile. Perhaps more promising is the
possibility of detecting DM annihilation from the centers of Galactic
subhalos, either from one of the known Milky Way dwarf satellite
galaxies or from a nearby dark clump \citep{Bergstroem1999, Baltz2000,
  CalcaneoRoldan2000, Tasitsiomi2002, Stoehr2003, Taylor2003,
  Evans2004, Aloisio2004, Koushiappas2004, Koushiappas2006,
  Diemand2007a, Strigari2007b, Pieri2008}. Lastly, an additional
source population might arise from local DM density enhancements
(mini-spikes) around black hole remnants of the first generation of
stars \citep{Bertone2005b}.

If GLAST detects photons originating in pair annihilations of DM
particles in the centers of subhalos, will it be possible to
distinguish this signal from conventional astrophysical gamma-ray
sources? While a definite answer to this question will have to await
actual GLAST data, \citet{Baltz2007} have argued that this should be
possible based on four criteria: (i) a hadronic spectrum from
monochromatic quarks, (ii) lack of time variability, (iii) spatial
extent, and (iv) a lack of emission at shorter wavelengths, except for
very diffuse inverse Compton and synchrotron radiation
\citep{Baltz2004,Colafrancesco2006,Colafrancesco2007}. Gamma-ray
pulsars are probably the most problematic astrophysical sources in
this regard, owing to the similarity of their spectra to a typical DM
annihilation spectrum. Fortunately they tend to lie in the Galactic
plane, often exhibit X-ray counterparts, and are point-like.

The observability of DM annihilation radiation originating in Galactic
subhalos depends on their abundance, distribution, and internal
structure, properties which must be determined by numerical
simulation, since they are the result of structure formation in the
highly non-linear regime. Previous investigations have typically only
indirectly made use of numerical simulations, employing
(semi-)analytic models that have been calibrated to published
numerical results. Often these results were derived from simulations
that do not resolve the relevant sub-galactic scales, and it is not
clear that a simple extrapolation is justified, since not all
substructure properties are scale invariant. Some past work
\citep{CalcaneoRoldan2000, Stoehr2003, Diemand2006, Athanassoula2008}
has directly used numerical simulations, but these studies either
suffered from insufficient resolution, didn't correct for effects from
the population of unresolved subhalos, or didn't realistically account
for the gamma-ray backgrounds against which individual subhalos must
compete.

Ground based gamma-ray detectors, i.e. atmospheric Cerenkov telescopes
(ACT) like H.E.S.S., VERITAS, MAGIC, and STACEE will provide
complementary information to GLAST's observations. However, owing to
the comparatively small field of view of ACT's, the detection of
individual subhalos with such observations would either have to rely
on serendipity to provide a detectable subhalo in the surveyed portion
of the sky, or they would have to target a known likely source such as
a nearby dark matter dominated dwarf galaxy \citep[e.g.][]{Albert2008,
  SanchezConde2007, Driscoll2007}. As advocated by
\citet{Koushiappas2004}, a better strategy might be to locate sources
with GLAST and then conduct follow-up observations with an ACT to
measure the gamma-ray spectrum out to higher energies.

Motivated by the imminent launch of the GLAST satellite, we present
here a comprehensive analysis of the detectability of individual DM
subhalos, taylored to GLAST expectations. Our method is based on a
hybrid approach, combining the highest resolution numerical
simulations of the distribution of Galactic DM substructure (described
in \S~\ref{sec:simulations}) with analytical models for the
extrapolation of the substructure hierarchy below the simulations'
resolution limit. We include a self-consistent treatment of subhalo
boost factors (\S~\ref{sec:boost}), motivated by our numerical
results, and a realistic treatment of the expected backgrounds that
individual subhalos must outshine (\S~\ref{sec:backgrounds}). We
consider a physically motivated region of particle physics parameter
space (\S~\ref{sec:DM_parameters}) and check that our results don't
violate existing EGRET constraints (\S~\ref{sec:egret_constraints}).
Our analysis results in quantitative predictions for the number of
observable subhalos as a function of particle physics parameters
(\S~\ref{sec:detectable}), and we also consider the possibility of
detecting microhalos below $1 \msun$ (\S~\ref{sec:convergence}). A
discussion of our results and conclusions can be found in
\S~\ref{sec:conclusion}.

\begin{figure*}[htp]
\begin{center}
\includegraphics[width=\textwidth]{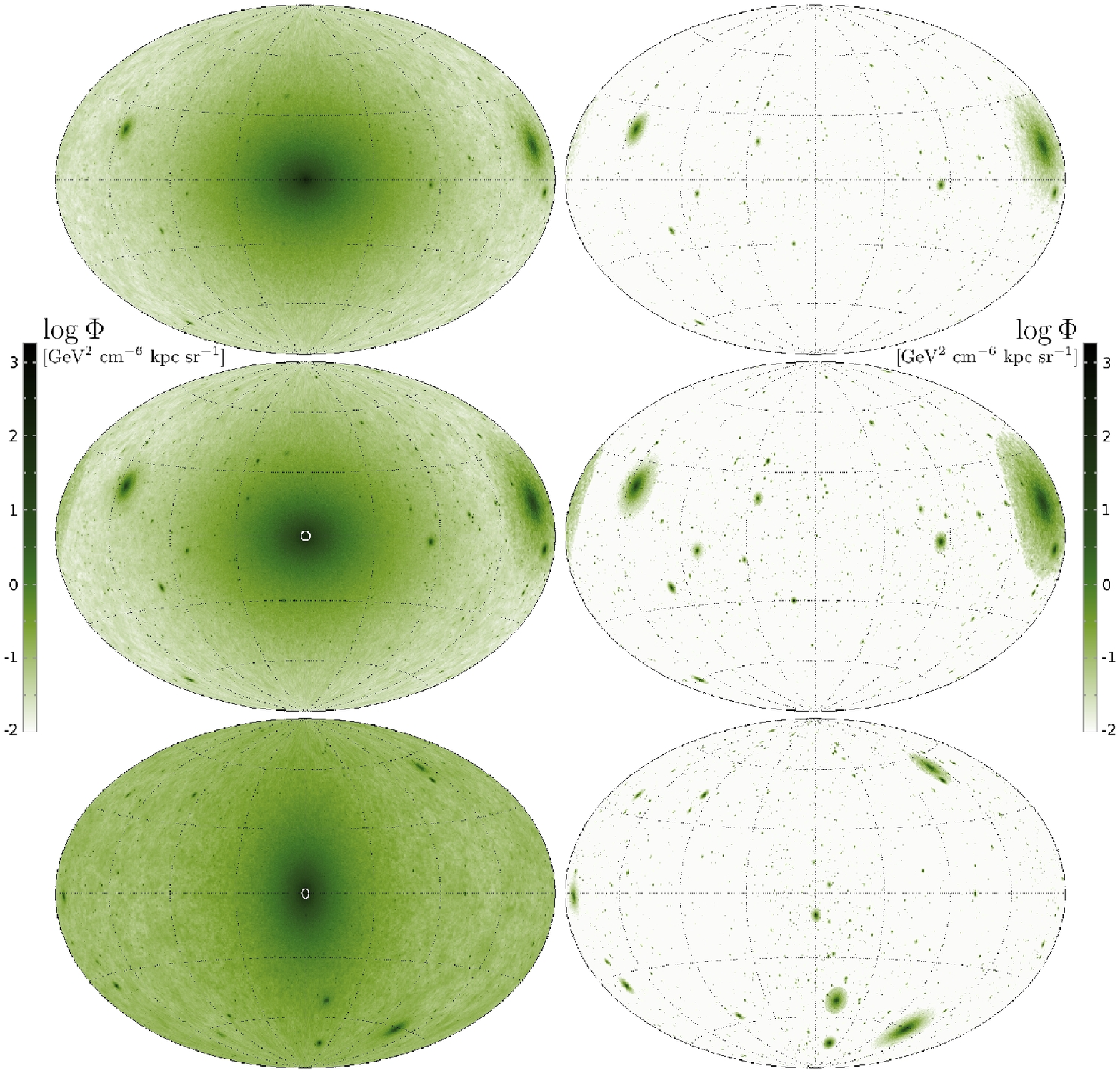}
\caption{Allsky maps of the annihilation signal from VL-II, for an
  observer 8 kpc from the halo center along the host halo's
  intermediate principal axis (top two rows) and along its major axis
  (bottom row). The maps in the left panel show the total annihilation
  signal from all DM particles within $\rtwo$, in the right panel only
  the signal from subhalo particles. In the top row we show the
  uncorrected signal directly from the simulation, in the bottom two
  rows the halo center, indicated with a small white ellipse, has been
  replaced with an artificial $\rho \propto r^{-1}$ cusp, and a
  mass-dependent boost factor (for $\mcut=10^{-6} \msun$,
  $\alpha=2.0$) has been applied to the subhalos. \label{fig:allsky}}
\end{center}
\end{figure*}

\section{Models and Methodology} \label{sec:models}


The number of DM annihilation gamma-ray photons from a solid angle
$\Delta \Omega$ along a given line of sight ($\theta$, $\phi$) over an
integration time of $\tau_{\rm exp}$ is given by
\begin{eqnarray}
N_\gamma(\theta,\phi) & & = \f{\Delta\Omega}{4\pi} \; \tau_{\rm exp}
\frac{\sigv}{2\MX^2} \left[ \int_{E_{\rm th}}^{\MX}
\left(\!\frac{dN_\gamma}{dE}\!\right) A_{\rm eff}(E) dE \right] \nonumber \\
&& \times \int_{\rm los} \! \rho(l)^2 dl.
\label{eq:annihilation}
\end{eqnarray}
This expression contains terms that depend a) on the properties of the
detector (the angular resolution $\Delta\Omega$, the energy dependent
effective area $A_{\rm eff}(E)$, the energy threshold $E_{\rm th}$,
and the total exposure time $\tau_{\rm exp}$), b) on the chosen
particle physics model describing the nature of the DM particle
(i.e. its mass $\MX$ and thermally averaged velocity-weighted
annihilation cross section $\sigv$, and the photon spectrum due to a
single annihilation event $dN_\gamma/dE$), and c) on the spatial
distribution of the DM within our halo. In this section we will focus
on the astrophysical contribution, while incorporating particle
physics and detector properties for the predictions of subhalo
detectability in the following section.

\begin{figure*}[htp]
\begin{center}
\includegraphics[width=\textwidth]{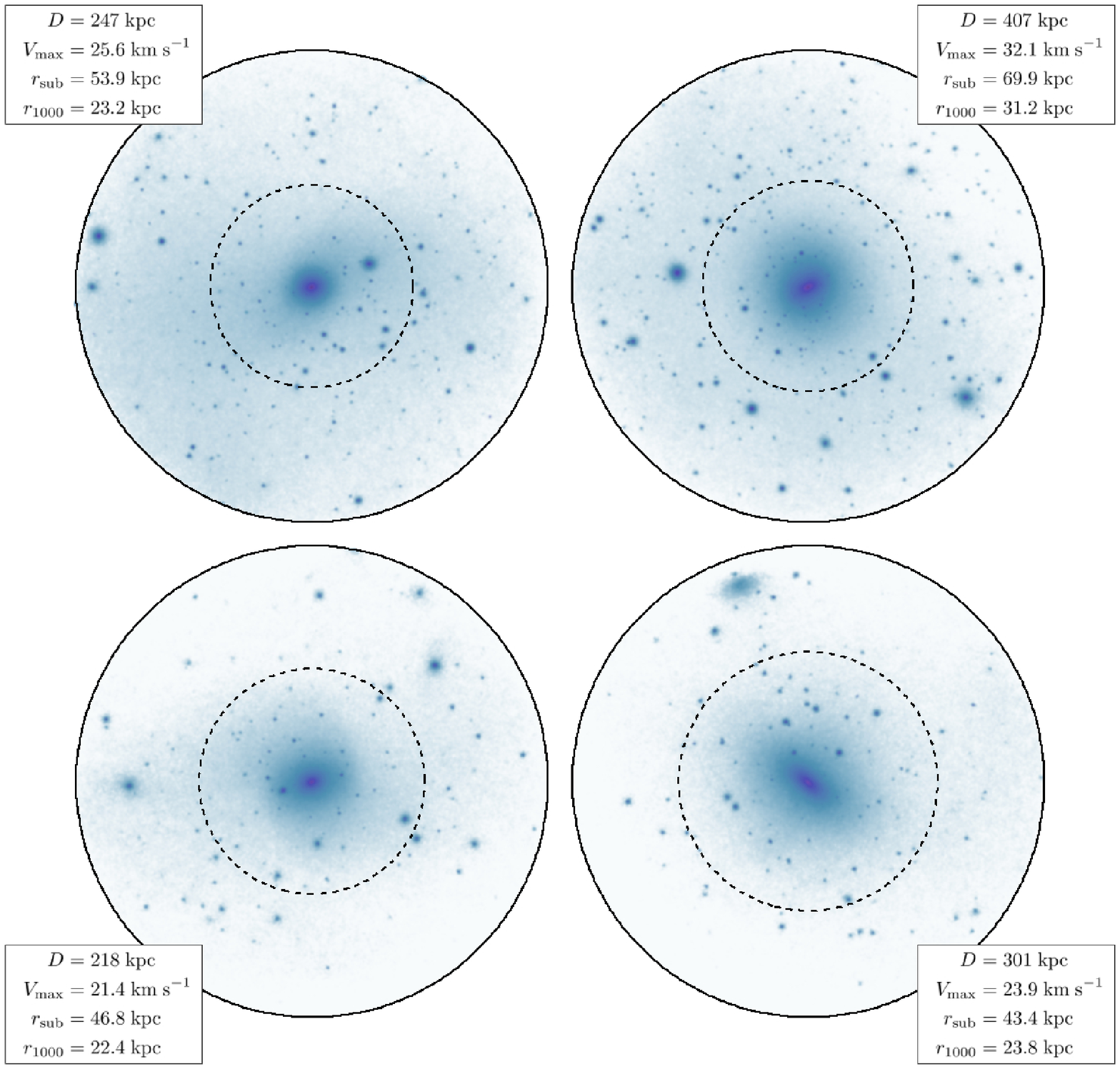}
\caption{Sub-substructure in four of VL-II's most massive subhalos.
  Shown are projections of $\rho^2$ for all particles within a
  subhalo's outer radius $r_{\rm sub}$. The dashed circle indicates
  the subhalo's $r_{1000}$. The clumpy sub-substructure boosts the
  total annihilation luminosity of its host subhalo.
  \label{fig:subhalos} }
\end{center}
\end{figure*}

\subsection{Simulations}\label{sec:simulations}

The ``Via Lactea'' simulations follow the formation and evolution of
the dark matter substructure of a Milky-Way-scale halo in a
cosmological volume. The first of these simulations (VL-I) consists of
234 million high resolution particles, covering the virial volume and
its immediate surroundings of a $M_{200}=1.77 \times 10^{12} \msun$
halo. With a particle mass of $\simeq 21,000 \msun$, this simulation
resolves around 10,000 subhalos within $\rtwo=388$ kpc\footnote{Note
  that we define $\rtwo$ as the radius enclosing 200 times the
  \textit{mean} density of the universe}. The global $z=0$ properties
of the host halo and the substructure population were presented in
\citet{Diemand2007a}, the joint temporal evolution of host
halo and substructure properties was discussed in \citet{Diemand2007b}, and an analysis of the shapes, alignment, and
spatial distribution of the subhalos can be found in
\citet{Kuhlen2007}. Recently we have completed ``Via Lactea II'', the
next generation of this series of simulations \citep{Diemand2008}.  At
a cost of over 1 million cpu-hours on Oak Ridge National Lab's
\textit{Jaguar} supercomputer, VL-II employs just over one billion
$4,100 \msun$ particles to model the formation of a $M_{200}=1.93
\times 10^{12} \msun$ halo and its substructure. For this simulation
we used for the first time a novel adaptive time-step method
\citep{Zemp2007}, which assigns time-steps equal to 1/16 of the local
dynamical time (but no shorter than 268,000 yr). This allows us to
accurately resolve the density structure much farther into the central
regions of the host halo. With VL-II we are able to resolve more than
50,000 individual subhalos within the host halo's $\rtwo=402$
kpc. Roughly 2,000 of these are found within 50 kpc of the center, and
20 within the solar circle. For reference the parameters of these two
simulations are summarized in Table~\ref{table:simulations}. The
calculations presented in this paper have been performed for both
simulations, but the final results are based on the higher resolution
VL-II.

Given the position of a fiducial observer, located 8 kpc from the host
halo center, we bin up the sky into a grid of $2400 \times 1200$
pixels, equally spaced in longitude ($\phi$) and in the cosine of the
co-latitude ($\cos \theta$). This pixel size was chosen to approximate
the best angular resolution of GLAST's LAT detector, equal to about 9
arcmin (see Section \ref{sec:detectability}). Each pixel corresponds
to a solid angle of $\Delta\Omega = (2\pi)/2400 \times 2/1200 = 4.363
\times 10^{-6}$ sr. We then calculate each particle's angular
coordinates on the sky $(\phi_i,\cos \theta_i)$, and sum up the
fluxes, $F_i = m_p \rho_i / 4 \pi d_i^2$, of all particles in a given
pixel. Here $d_i$ is the distance from the $i^{\rm th}$ particle to
the observer, $\rho_i$ is its density, measured over a 32 particle
SPH-kernel, and we have approximated $\int \rho_i^2 dV$ as $m_p
\rho_i$, which is appropriate for a collection of discrete simulated
DM particles of mass $m_p$. In order to minimize shot noise arising
from particle discreteness we smooth each particle's contribution
using a projected SPH-kernel of angular width $h_i/d_i$, where $h_i$
is half the radius encompassing all 32 neighboring particles.  The
resulting map is presented in the top left panel of
Fig.~\ref{fig:allsky} in units of GeV$^2$ kpc cm$^{-6}$ sr$^{-1}$ and
depends only the choice of the observer's location and the dark matter
distribution. The flux from only those particles belonging to subhalos
is shown in the top right panel.

\subsection{Subhalo Boost Factor}\label{sec:boost}

The hierarchical nature of CDM structure formation implies that
substructure should be expected not only in the host halo, but also in
individual subhalos. Indeed, in VL-II we are able to resolve some of
these sub-subhalos in the most massive of our subhalos.  In
\citet{Diemand2008} we show that the mean abundance of
sub-substructure is consistent with a scaled-down version of the VL-II
host halo, i.e. once the sub-subhalos' $\Vmax$'s are scaled by the
$\Vmax$ of \textit{their} host subhalo, the cumulative
$\Vmax$-function ($N(>\Vmax)$) agrees with the overall subhalo
$\Vmax$-function. A density squared projection of four of the most
massive subhalos is shown in Fig.~\ref{fig:subhalos}. Only particles
within the subhalo's outer radius $r_{\rm sub}$ have been included in
the projection.\footnote{$r_{\rm sub}$ is defined here as the radius
  at which the total density becomes roughly constant.} The subhalos'
$r_{1000}$, the radius at which the mean enclosed density is equal to
1000 times the mean matter density of the universe, is overplotted
with a dashed line.

Even with VL-II's impressive dynamical range of $\sim 10^6$ in mass,
however, we still resolve only a small fraction of the total expected
DM substructure hierarchy. In principle this hierarchy could extend
all the way down to the cut-off in the matter power spectrum, set by
collisional damping and free streaming in the early universe
\citep{Green2005, Loeb2005}. For WIMP dark matter, typical kinetic
decoupling temperatures range from MeV to GeV, corresponding to
cut-off masses of $\mcut = 10^{-12}$ to $10^{-4} \msun$
\citep{Profumo2006}, some 10 to 20 orders of magnitude below VL-II's
mass resolution. Since the annihilation rate goes as density squared,
any clumpiness will lead to an enhancement of the total luminosity
compared to a smooth mass distribution.

\begin{figure}
\begin{center}
\includegraphics[width=\columnwidth]{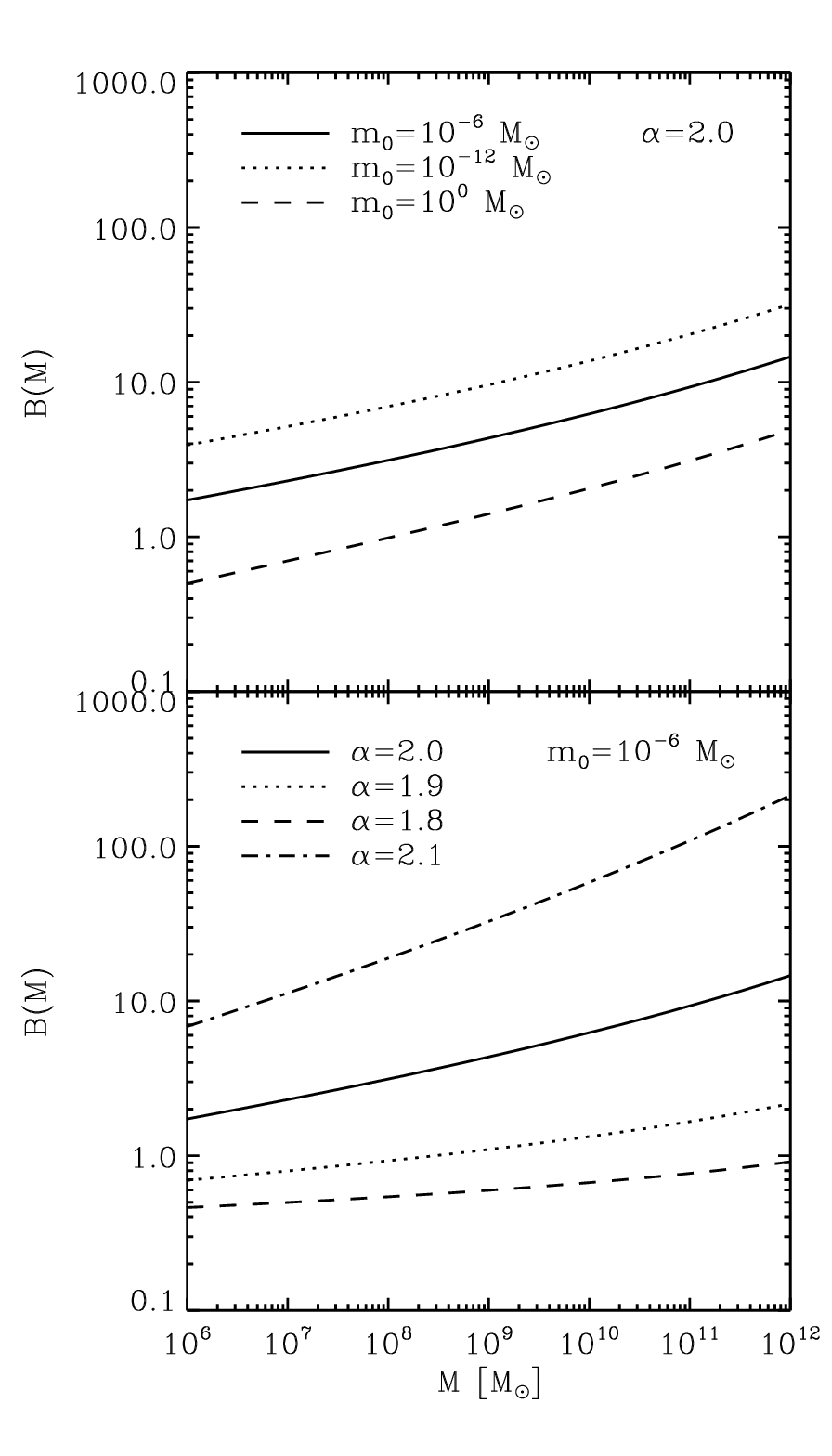}
\caption{The annihilation luminosity boost factor due to substructure
  below VL-II's resolution limit versus subhalo mass, for different
  subhalo mass functions. \textit{Top panel:} Dependence on the cutoff
  mass $\mcut$ for slope $\alpha=2.0$. \textit{Bottom panel:}
  Dependence on $\alpha$ for $\mcut=10^{-6} \msun$.\label{fig:boost} }
\end{center}
\end{figure}

The unresolved portion of the substructure hierarchy will thus lead to
a boost in the true annihilation luminosity of individual subhalos,
compared to the value determined directly from the numerical
simulation. The magnitude of this boost will depend sensitively on the
properties of the subhalo population, in particular on the slope and
low mass cut-off of the subhalo mass function and on the subhalo
concentration-mass relation. Our analytic boost factor model is based
on the one presented in \citet{Strigari2007a}. The true luminosity of a
subhalo of mass M is given by
\begin{equation}
L(M) = \left[ 1 + B(M) \right] \tilde{L}(M),
\end{equation}
where
\begin{equation}
\tilde{L}(M) \propto \int_{V} \rho^2_{\rm sub} dV = \sum_{r_i<\rsub} \rho_i m_p
\end{equation}
is the luminosity of the subhalo determined from all simulation
particles within the subhalo's outer radius and $\rho_i$ is the
density of the $i^{\rm th}$ such particle. A subhalo's $B(M)$ can be
calculated by integrating luminosities over its own sub-subhalo
population:
\begin{eqnarray}
B(M) & = & \f{1}{\tilde{L}(M)} \int_{\mcut}^{m_1} \f{dN}{dm} L(m) dm \\
     & = & \f{1}{\tilde{L}(M)} \int_{\mcut}^{m_1} \f{dN}{dm} \left[1 + B(m) \right] \tilde{L}(m) dm. \label{eq:boost1}
\end{eqnarray}
Here $dN/dm$ is the sub-subhalo mass function, and the integration
extends from $\mcut$, the low mass cut-off of the substructure
hierarchy, to an upper limit of $m_1={\rm min}\{10^6 \msun, 0.1M\}$,
such that only substructure below VL-II's resolution limit of $\sim
10^6 \msun$ contribute. For subhalos below $10^7 \msun$ we cap the
integration at $0.1M$ under the assumption that efficient dynamical
friction would have lead to the tidal destruction of larger
sub-subhalos. For a power law substructure mass function $dN/dm = A/M
(m/M)^{-\alpha}$, Eq.~\ref{eq:boost1} becomes
\begin{equation}
B(M) = \f{A}{\tilde{L}(M)} \int_{\ln m_0}^{\ln m_1} \left( \f{m}{M} \right)^{1-\alpha} \left[ 1 + B(m) \right] \tilde{L}(m) d\!\ln m. \label{eq:boost2}
\end{equation}
Motivated by our numerical simulations \citep{Diemand2004a,
  Diemand2007a} and semi-analytic studies \citep{Zentner2003}, we
normalize the sub-subhalo mass function by setting the mass fraction
in subclumps with masses $10^{-5} < m/M < 10^{-2}$ equal to 10\%.

For the determination of $\tilde{L}(M)$ we have assumed an NFW density
profile, in which case the total annihilation luminosity of a halo of
mass $M$ and concentration $c=\rvir/r_s$ is given by
\begin{equation}
\tilde{L}(M,c) \propto \rho_s^2 r_s^3 \propto M \f{c^3}{f(c)^2},
\end{equation}
where $f(c) = \ln (1+c) - c/(1+c)$. We use the \citet{Bullock2001}
concentration-mass relation for field halos, albeit with a somewhat
smaller value of the normalization, $K=3.75$ \citep[as suggested
by][]{Kuhlen2005,Maccio2007}. For the cosmology used in the VL
simulations and halos masses between $10^6$ and $10^{10} \msun$, the
c(M) relation is approximately $c(M) \approx 18 (M/10^8
\msun)^{-0.06}$, which corresponds to $\tilde{L}(M) \propto M^{0.87}$,
i.e. the annihilation luminosity scales almost linearly with mass, in
agreement with results from numerical simulations \citep{Stoehr2003,
  Diemand2007a}. Note that in our numerical simulations we find
systematically higher subhalo concentrations closer to the host halo
center. This trend does not affect the magnitude of the boost factor,
but translates to a radial trend in subhalo luminosity (see Section
3.1).

Eq.~\ref{eq:boost2} is solved numerically using the boundary condition
$B(\mcut)=0$. The resulting relation is plotted in
Fig.~\ref{fig:boost}, for $\alpha=2.0$ and different values of $\mcut$
in the top panel, and for $\mcut=10^{-6}\msun$ and different values of
$\alpha$ in the bottom panel. Overall we find relatively modest boost
factors on the order of a few, ranging up to $\sim 10$ for the most
massive subhalos. Generally more massive halos have larger boost
factors, simply because their subhalo population covers more of the
total subhalo hierarchy. For the same reason, smaller values of
$\mcut$ lead to larger boost factors. For $\alpha<2.0$ $B(M)$ has a
weaker mass dependence and is less sensitive to $\mcut$, since in this
case more massive halos are relatively more important. Our results are
in agreement with the analytic upper limits of \citet{Strigari2007a}
and the recent calculations of \citet{Lavalle2008}.

A fit to the cumulative subhalo mass function in our simulations is
consistent with $\alpha=2$ \citep{Diemand2007a}, which implies equal
mass in subhalos per decade of subhalo mass. However, fits to the
differential mass function tend to favor slightly shallower slopes of
$1.8-1.9$ \citep{Stoehr2003, Madau2008}, possibly because they are
more sensitive to the lower mass end, where resolution effects may
artificially flatten the slope. In this work we use $\alpha=2.0$ and
$\mcut=10^6 \msun$ as our fiducial model, but present results for a
range of different $\alpha$ and $\mcut$.

\subsection{Central Flux Corrections}\label{sec:central_flux_corrections}

The host halo center is another area where our simulation must be
corrected to account for the artificially low density caused by the
finite numerical resolution \citep{Diemand2004b}. Based on numerical
convergence studies \citep{Diemand2005a} we believe that we can trust
the radial density profile of the VL-I host halo down to $r_{\rm
  conv}=3.4 \times 10^{-3} \rtwo = 1.3$ kpc \citep{Diemand2007a},
corresponding to about 10$^\circ$ from the center. The higher mass
resolution and improved time-step criterion in VL-II results in a much
smaller convergence radius of $r_{\rm conv}=380$ pc. The flux derived
directly from the simulated particles in VL-II will thus only
underestimate the true annihilation flux within the inner $\sim
2^\circ$ from the center. An additional uncertainty arises from the
fact that our purely collisionless DM simulation completely neglect
the effect of baryons.  While this is not a problem for the signal
from individual subhalos, which are small enough that baryonic effects
are likely negligible, the central region of our host halo most likely
would have been affected by gas cooling, star formation, and stellar
dynamical processes. It is not immediately obvious how such baryonic
effects would alter the central DM distribution. Adiabatic contraction
\citep{Blumenthal1986,Gnedin2004a} would lead to a steepening of the
central DM density profile at scales of a few kpc and below. A recent
study of scaling relations in spiral galaxies, however, seems to favor
models of spiral galaxy formation without adiabatic contraction, and
suggests that clumpy gas accretion might have reduced central DM
densities \citep{Dutton2007}. Stirring by a stellar bar could also
eject DM from the central regions \citep[and references
therein]{Weinberg2007}. On much smaller scales (central few pc), the
presence of a supermassive black hole (SMBH) would lead to the
creation of an $r^{-1.5}$ mini-cusp \citep{Gnedin2004b}. A SMBH
binary, on the other hand, could have removed DM from the very center
\citep{Merritt2002}.

\begin{figure}
\begin{center}
\includegraphics[width=\columnwidth]{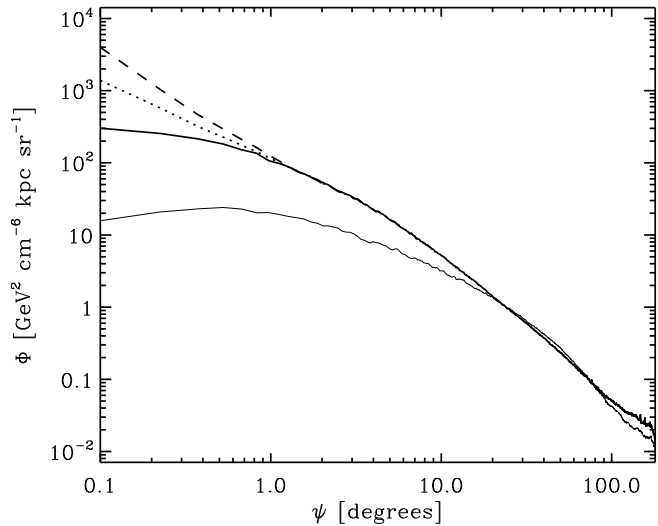}
\caption{The gamma-ray intensity from DM annihilations in the smooth
  host halo versus the angle from the Galactic center, directly from
  the simulated VL-II particles (\textit{thick solid line}), and with
  an artificial central cusp with central slope equal to 1.0
  (\textit{dotted line}) and 1.2 (\textit{dashed line}). For
  comparison the smooth host halo flux from the VL-I simulation is
  overplotted (\textit{lower thin solid line}). \label{fig:host_flux}
}
\end{center}
\end{figure}

Here we have attempted to correct for the artificially low central
density by excising all simulated particles within $r_{\rm conv}$ and
replacing them with an artificial cusp consisting of about 25 million
5 $\msun$ particles, distributed in an ellipsoid matched to the shape
and density of the VL-II host halo at $r_{\rm conv}$, and extended
inwards with a power law density profile. The resulting radial flux
profile (excluding subhalos) is plotted in Fig.~\ref{fig:host_flux},
for the uncorrected case and for central density slopes of -1 and
-1.2. 

Note that a similar correction should be applied to the centers of all
the subhalos as well. This, however, is computationally very
expensive, and we omit a direct correction of the particle
distribution in favor of applying an \textit{a posteriori} correction
to the central pixel of each subhalo in the final allsky map. For each
subhalo we estimate the central surface brightness according to
\begin{equation}
\Phi_c = \f{1}{\Delta\Omega} \mathcal{F}_{\rm halo}(\rho_s,r_s,D),
\end{equation}
where $\mathcal{F}_{\rm halo}$ is the flux from the central region
subtended by solid angle $\Delta\Omega$ of a subhalo with NFW scale
density $\rho_s$ and scale radius $r_s$ at a distance $D$ (see
Eq.~\ref{eq:Flux_halo}). We estimate $\rho_s$ and $r_s$ from the
measured values of $\Vmax$ and $\rVmax$ according to the relations
\begin{eqnarray}
r_s & = & \f{\rVmax}{2.163} \\
\rho_s & = & \f{4.625}{4 \pi G} \left( \f{\Vmax}{r_s} \right)^2,
\end{eqnarray}
which hold for an NFW density profile. If the central pixel of a
subhalo has $\Phi<\Phi_c$, we set it equal to $\Phi_c$. Note that we
first apply the boost factor correction to all particles of a given
subhalo, and then apply this flux correction to the central pixel. We
only correct the central pixel, so subhalos for which $r_s$ subtends
more than one pixel (1.7\% of all VL-II subhalos) will not be fully
corrected.

\subsection{Corrected Allsky Maps}

In Fig.~\ref{fig:allsky} we show both unmodified and corrected DM
annihilation flux allsky maps. While the topmost row shows the signal
as calculated directly from the numerical simulation, in the center
row the subhalo particle fluxes have been boosted by $B(M)$, assuming
$(\alpha,\mcut)=(2.0,10^{-6})$, the central pixel flux correction has
been applied, and the central region of the host halo has been
replaced with an artificial density cusp of slope -1. In these two
rows the observer is located at 8 kpc along the intermediate axis of
the host halo ellipsoid. In the panels on the right hand side, we show
only the fluxes from particles belonging to subhalos, i.e. within a
given subhalo's outer radius. The addition of a substructure boost
factor lifts the brightness of a number of subhalos over the level of
the diffuse host halo flux. The bottom panel shows the corrected case
for an observer placed at 8 kpc along the major axis of the host halo.
In this case the observer is deeper inside the host halo mass
distribution, and the resulting diffuse host halo flux dominates over
most of the subhalos.

\section{Subhalo Detectability} \label{sec:detectability}

In order to make predictions for the number of detectable subhalos, we
must convert the simulation fluxes presented in
Section~\ref{sec:models} into actual gamma-ray photon fluxes and
compare these to the expected background signal. Following previous
work \citep{Stoehr2003}, we define for each subhalo in our allsky map
a ``detection significance''
\begin{equation}
\sig = \f{\sum_i N_{s,i}}{\sqrt{\sum_i N_{b,i}}},
\end{equation}
where $N_{s,i}$ and $N_{b,i}$ are the number of source and background
gamma-rays received on a given pixel, and the sums are over all pixels
with signal-to-noise ($N_{s,i}/N_{b,i}^{1/2}$) greater than unity
covered by a subhalo. For this purpose we must make assumptions about
the observation (the detector's effective area and angular resolution
as a function of energy, the exposure time and energy threshold), the
astrophysical backgrounds (extragalactic and Galactic), and, most
importantly, the nature of the DM particle (its mass and annihilation
cross section, the annihilation spectrum).

Here we adopt parameters appropriate for a GLAST observation with the
Large Area Telescope (LAT): an exposure time of $\tau_{\rm exp}=2$
years, the expected energy-dependent effective area $A_{\rm eff}(E)$,
and an energy threshold of $E_{\rm th}=3.0$ GeV, above which the
flux-weighted mean angular resolution
\begin{equation}
  \langle \Delta \theta \rangle \equiv \f{\int_{E_{\rm th}}^\infty \Delta\theta(E) \; A_{\rm eff}(E) \; (dN_{\gamma}/dE) \; d\!E}{\int_{E_{\rm th}}^\infty A_{\rm eff}(E) \; (dN_{\gamma}/dE) \; d\!E} 
\end{equation}
is equal to 9 arcmin. The functional form of $A_{\rm eff}(E)$ and
$\Delta\theta(E)$ was obtained from the GLAST LAT Performance
webpage\footnote{http://www-glast.slac.stanford.edu/software/IS/glast\_lat\_\\performance.htm}.

\subsection{Backgrounds}\label{sec:backgrounds}

The annihilation signal from individual extended subhalos must compete
with a number of diffuse gamma-ray backgrounds, of both astrophysical
and particle physics origin. The Energetic Gamma Ray Experiment
Telescope (EGRET) aboard the Compton Gamma Ray Observatory satellite
conducted a gamma-ray allsky survey in the 1990's, detecting a diffuse
background consisting of an isotropic extragalactic
\citep{Sreekumar1998} as well as a Galactic component
\citep{Hunter1997}. The extragalactic component is probably dominated
by unresolved blazars \citep{Stecker1996,Giommi2006}, although
star-bursting galaxies and galaxy clusters might add a substantial
contribution \citep{Chiang1998,Dermer2007}. The spectrum of this
extragalactic background is well described by a power-law photon
spectrum in the 30 MeV to 100 GeV energy range \citep{Sreekumar1998}:
\begin{eqnarray}
\Phi_{\rm eg} & = &(7.32 \pm 0.34) \times 10^{-9} \quad {\rm MeV}^{-1} {\rm cm}^{-2} {\rm s}^{-1} {\rm sr}^{-1} \nonumber \\
 & & \left( \f{E}{451{\rm MeV}} \right)^{(-2.10 \pm 0.03)}.
\label{eq:EGRET}
\end{eqnarray}

The diffuse Galactic gamma-ray background is produced by energetic
interactions of cosmic ray nucleons with interstellar medium (ISM)
atoms through neutral pion production, as well as from energetic
electrons by inverse Compton upscattering of lower energy photons and
bremsstrahlung. Modeling this background under the assumption that the
locally measured electron and proton energy spectra are representative
of the Galaxy as a whole (``conventional model'') underpredicts the
flux in the measured EGRET spectrum above 1 GeV by about a factor of
two \citep{Hunter1997}. \citet{Strong2004a} provide an explanation for
this ``GeV excess'' by modeling cosmic ray propagation within the
Galaxy, allowing for mild departures in the electron and nucleon
spectra from the conventional model, and using a cosmic ray source
distribution based on the observed distribution of pulsars and
supernova remnants in conjunction with a variable CO-to-H$_2$
conversion factor \citep{Strong2004b}. This ``optimized model''
emphasizes the importance of inverse Compton emission and is able to
reproduce the EGRET observations in all directions of the sky. Here we
have employed the GALPROP\footnote{available from
\texttt{http://galprop.stanford.edu/web\_galprop/\\galprop\_home.html}}
(v50p) cosmic ray propagation code \citep[][and references
therein]{Strong1998,Moskalenko2002} to calculate an ``optimized
model'' allsky map of the gamma ray emissivity due to cosmic ray
interactions with the Galactic ISM. We have limited the pixel size of
this map to 0.5$^\circ$, corresponding to the angular resolution of
the HI and CO surveys used in GALPROP.

\begin{figure}
\begin{center}
\includegraphics[width=\columnwidth]{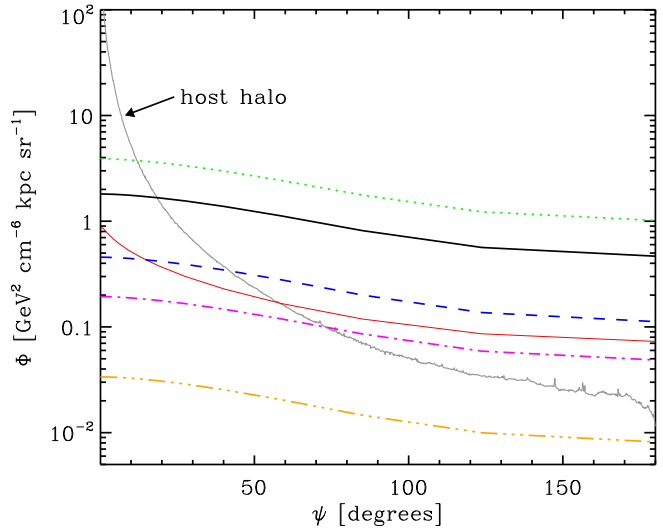}
\caption{Diffuse flux due to undetectable subhalos as a function of
  angle $\psi$ from the Galactic center, for a number of different
  subhalo mass functions. The thick lines show models with an
  anti-biased radial distribution, concentrations increasing towards
  the host center, and different values of the mass function slope
  $\alpha$ and low mass cutoff $\mcut$:
  $(\alpha,\mcut/\msun)=(2.0,10^{-6})$ (\textit{solid}),
  $(2.0,10^{-12})$ (\textit{dotted}), $(2.0,1)$ (\textit{dashed}),
  $(1.9,10^{-6})$ (\textit{dot-dashed}), $(1.8,10^{-6})$
  (\textit{triple-dot-dashed}).  The thin solid line represents the
  original \citet{Pieri2008} model ($B_{\rm ref,z0}$), with $\alpha=2.0$,
  $\mcut=10^{-6}\msun$, an un-biased radial distribution, and no
  radial concentration dependence. The flux from the smooth host halo
  is overplotted with the grey line, see Fig.~\ref{fig:host_flux}.
  \label{fig:diffuse} }
\end{center}
\end{figure}

In addition to these astrophysical backgrounds, individual detectable
subhalos must outshine DM annihilations from the smooth host halo as
well as the background from the population of individually
undetectable subhalos \citep{Pieri2008}. Note that these undetectable
subhalos do not simply uniformly boost the flux from the smooth host
halo. Tidal disruption of satellites is more effective close to the
host halo center, leading to a spatial distribution of subhalos that
is antibiased with respect to the host halo mass distribution
\citep{Kuhlen2007, Madau2008}. The resulting background will have a
shallower angular dependence than the smooth host halo signal, and can
dominate it at large angular distances from the center.

To determine the magnitude and angular dependence of this background
we repeat the calculation presented in \citet{Pieri2008}, with three
important differences.  Firstly, we use an antibiased subhalo spatial
distribution for which $n_{\rm sub}(r)/\rho_{\rm host}(r) \propto r$
\citep[see][]{Kuhlen2007}, as opposed to one that follows the host
halo mass distribution down to some hard cut-off $r_{\rm min}(M)$ as
in \citet{Pieri2008}. Secondly, we allow a range of values of the
subhalo mass function slope $\alpha$ and cutoff mass $\mcut$. This can
make a big difference, since by number the population of individually
undetected subhalos is dominated by objects with masses close to
$\mcut$. Lastly, we include a radial dependence in the subhalo
mass-concentration relation, motivated by numerical simulations which
tend to find higher concentrations for subhalos closer to the host
halo center \citep{Diemand2007b,Diemand2008},
\begin{equation}
c_0^{\rm sub}(M,R) = c_0^{\rm B01}(M) \left( \f{R}{\rtwo} \right)^{-0.286} \hspace{-0.35in},\label{eq:c_of_r}
\end{equation}
where $c_0^{\rm B01}(M)$ is the median concentration of subhalo of
mass $M$, as given by the \citet{Bullock2001} model for field halos.
With this scaling, subhalos at $R_\odot$ are three times as
concentrated as field halos. We also include a log-normal scatter
around this median, with width $\sigma_{\log_{10} c}=0.14$
\citep{Wechsler2002}. In Fig.~\ref{fig:diffuse} we present the
resulting background flux as a function of angle from the halo center,
and also show the effects of our modifications on the original
\citet{Pieri2008} prescription (using their $B_{\rm ref, z0}$ model).
A more detailed explanation of our calculation of this background is
included in the Appendix~\ref{sec:AppendixA}.

We find that our use of an anti-biased radial distribution leads to a
diffuse subhalo flux that is almost independent of the viewing
direction. The median galacto-centric distance of a subhalo (i.e. the
radius enclosing half of all subhalos) is about 200 kpc in the
anti-biased case, but only 100 kpc for the unbiased distribution used
by \citet{Pieri2008}. The fraction of subhalos within 8 kpc (within
$r_s^{\rm VL-II}=21$ kpc) is $7 \times 10^{-4}$ (0.01) in the anti-biased
case, and 0.02 (0.1) for the unbiased distribution. In the unbiased
case, subhalos within 8 kpc of the Galactic center contribute about
90\% of the subhalo diffuse flux towards the Galactic center, whereas
they only make up 40\% of the flux in the anti-biased case. The shift
towards larger distances also leads to an overall reduction in the
amplitude of the flux.

\begin{table}
\caption{Subhalo Mass Function Models}
\begin{tabular}{ccccccc}
\hline
\hline
$\alpha$ & $\mcut$ & $N_{\rm tot}$ & $M_{\rm tot}$ & $f_{\rm tot}$ & $M_{\rm u}$ & $f_{\rm u}$ \\
 & $(\msun)$ & & $(\msun)$ & & $(\msun)$ & \\
\hline
   2.0   & $10^{-6}$ & $2.5 \times 10^{16}$ & $9.3 \times 10^{11}$ & 0.53          & $7.0 \times 10^{11}$ & 0.40 \\
   1.9   & $10^{-6}$ & $9.2 \times 10^{14}$ & $3.2 \times 10^{11}$ & 0.19          & $1.2 \times 10^{11}$ & 0.070 \\
   1.8   & $10^{-6}$ & $3.3 \times 10^{13}$ & $2.1 \times 10^{11}$ & 0.12          & $3.3 \times 10^{10}$ & 0.018 \\
   2.0   & $1$ & $2.5 \times 10^{10}$ & $5.8 \times 10^{11}$ & 0.33          & $3.5 \times 10^{11}$ & 0.20 \\
   2.0   & $10^{-12}$ & $2.5 \times 10^{22}$ & $1.3 \times 10^{12}$ & 0.73          & $1.0 \times 10^{12}$ & 0.60 \\
\end{tabular}
\tablecomments{The total number ($N_{\rm tot}$), mass ($M_{\rm tot}$),
  and mass fraction ($f_{\rm tot}=M_{\rm tot}/\mvir$) by extrapolation
  of the subhalo mass function with slope $\alpha$ and cutoff $\mcut$,
  normalized to give $f=0.1$ in the interval
  $10^{-5}<M/\mvir<10^{-2}$. $M_{\rm u}$ and $f_{\rm u}$ are the mass
  and mass fraction of all subhalos below VL-II's resolution limit of
  $\sim 10^6\msun$.}
\label{table:mf_models}
\end{table}

The final background considered here is due to annihilations from the
smooth host halo. For this component we simply use the angular flux
distribution calculated from all simulated particles that don't belong
to any subhalos. Since higher numerical resolution would have resolved
some of this DM mass into individual subhalos, whose contribution we
have accounted for above, we uniformly reduce the smooth halo flux by
a factor $(1-f_u)$, where $f_u$ is the mass fraction below $10^6
\msun$ (last column in Table~\ref{table:mf_models}).

\subsection{Particle Physics Parameters}\label{sec:DM_parameters}

The particle physics dependence of the annihilation signal
(Eq.~\ref{eq:annihilation}) enters through three factors: $\MX$, the
mass of the DM particle, $\sigv$, the thermally averaged
velocity-weighted annihilation cross section, and $dN_\gamma/dE$, the
photon spectrum resulting from a single annihilation event. The
physical nature of DM is currently unknown, and a plethora of particle
physics models have been proposed to explain its existence. It should
be noted that not all of these models result in a DM particle capable
of annihilating, but those models are not of interest for the present
work. Instead we consider here only the class of models in which the
DM is a weakly interacting massive particle (WIMP), such as the
neutralino in supersymmetric extensions of the standard model or
Kaluza-Klein excitations of standard model fields in models with
universal extra dimensions \citep[for a recent review of particle DM
  theories, see][]{Bertone2005a}.

For any given class of model it is possible to determine a range of
$\MX$ and $\sigv$ resulting in a current relic DM density that is
consistent with the WMAP measurement of $\Omega_\chi
h^2=0.1105^{+0.0039}_{-0.0038}$ \citep{Spergel2007}. Typical values
for $\MX$ are 50 GeV up to $\sim 1$ TeV, and a simple estimate of the
cross section is $\sigv = 3 \times 10^{-27} {\rm cm}^3 {\rm s}^{-1} /
\Omega_\chi h^2 \approx 3 \times 10^{-26}$ cm$^3$
s$^{-1}$\citep{Jungman1996}. However, this naive relation can fail
badly \citep{Profumo2005}, and a much wider range of cross sections,
up to $\sigv \sim 10^{-24}$ cm$^3$ s$^{-1}$ for $\MX<200$ GeV
\citep[e.g. Fig.17 in][]{Colafrancesco2006}, should be considered
viable. In this work we consider values of $\MX$ from 50 to 500 GeV,
and $\sigv$ from $10^{-26}$ to $10^{-25}$ cm$^3$ s$^{-1}$.

\begin{figure}
\begin{center}
\includegraphics[width=\columnwidth]{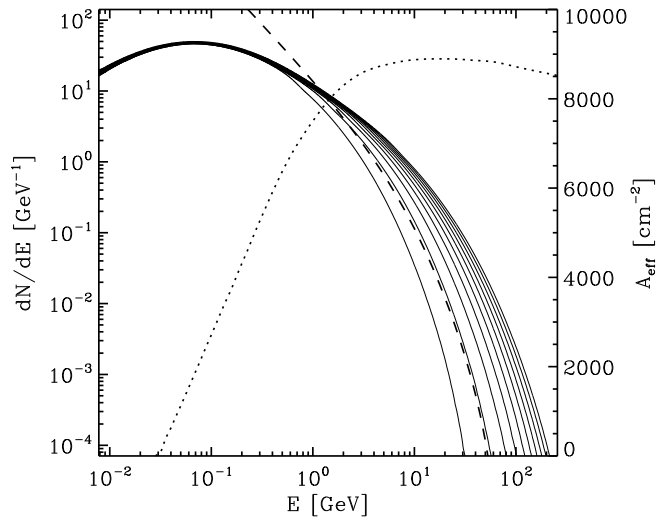}
\caption{The gamma-ray spectrum $dN/dE$ resulting from WIMP
  annihilations into $b\bar{b}$ quark pairs, calculated with
  \texttt{DarkSUSY} \citep{Gondolo2004}, for $\MX$ between 50 and 500
  GeV in increments of 50 GeV. The dashed line shows the
  \citet{Fornengo2004} fitting function for $\MX=100$ GeV and the
  dotted line the expected effective area of the GLAST/LAT detector
  (right abscissa). \label{fig:spectrum} }
\end{center}
\end{figure}

WIMP DM particles can annihilate into a range of different particle
pairs, including quarks, leptons (e.g. $\tau$'s), gauge bosons ($Z^0$,
$W^\pm$), gluons, and Higgs particles. The subsequent decay and
hadronization of the annihilation products leads to a spectrum of
gamma-rays, mostly resulting from the decay of $\pi^0$ mesons. The
shape of this spectrum is almost independent of the annihilation
branch \citep[with the exception of the $\tau$ branch, which has a
harder spectrum,][]{Cesarini2004,Fornengo2004}, and can be calculated
using Monte-Carlo simulation, for example with the \texttt{PYTHIA}
package \citep{Sjoestrand2001}. Above $x=(E_\gamma/\MX)=0.01$ the
spectrum is typically well fit by a function of the form $\alpha_0
x^{3/2} \exp(-\alpha_1 x)$, with $(\alpha_0,\alpha_1)$ dependent on
the final state \citep{Bergstroem1998,Koushiappas2004,Fornengo2004}.
GLAST has non-negligible sensitivity down to $\sim 0.1$GeV ($x<0.01$),
where the spectrum has turned over and the $x^{3/2}$ dependence is no
longer a good match. Here we calculate $dN/dE$ assuming a 100\%
branching ratio into $b\bar{b}$ quarks \citep{Baltz2007,Pieri2008} and
calculate the spectrum directly using the \texttt{DarkSUSY} code
\citep{Gondolo2004}. Annihilation into two photons or a photon and a
$Z^0$, resulting in a monochromatic gamma-ray line signal, is also
possible, but is one-loop suppressed \citep{Bergstroem1997} and we do
not include such channels here. Fig.~\ref{fig:spectrum} shows the
$b\bar{b}$ spectrum for $\MX$ between 50 and 200 GeV, together with
the \citet{Fornengo2004} fitting function and the GLAST/LAT $A_{\rm
  eff}(E)$. For $\MX=100$ GeV, this spectrum results in 32.6, 13.4,
and 4.58 gamma-rays per annihilation above 0.1, 1.0, and 3.0 GeV,
respectively.

Recently \citet{Bringmann2008} presented new calculations of the
gamma-ray spectrum from DM annihilation, including electromagnetic
radiative corrections to all leading annihilation processes. In some
cases the inclusion of internal bremsstrahlung (IB) photons leads to
large enhancements of the gamma-ray fluxes, and also sharpens the
spectral feature at the mass of the DM particle. Above energies of
$0.6 \MX$, IB photons typically increase the gamma-ray flux by no more
than a factor of two over the range of $\MX$ considered here. For the
much lower energy threshold of 3 GeV employed in this study, the
enhancement from IB photons is negligible, and thus we have not
included IB photons here.

\begin{figure*}
\begin{center}
\includegraphics[width=\textwidth]{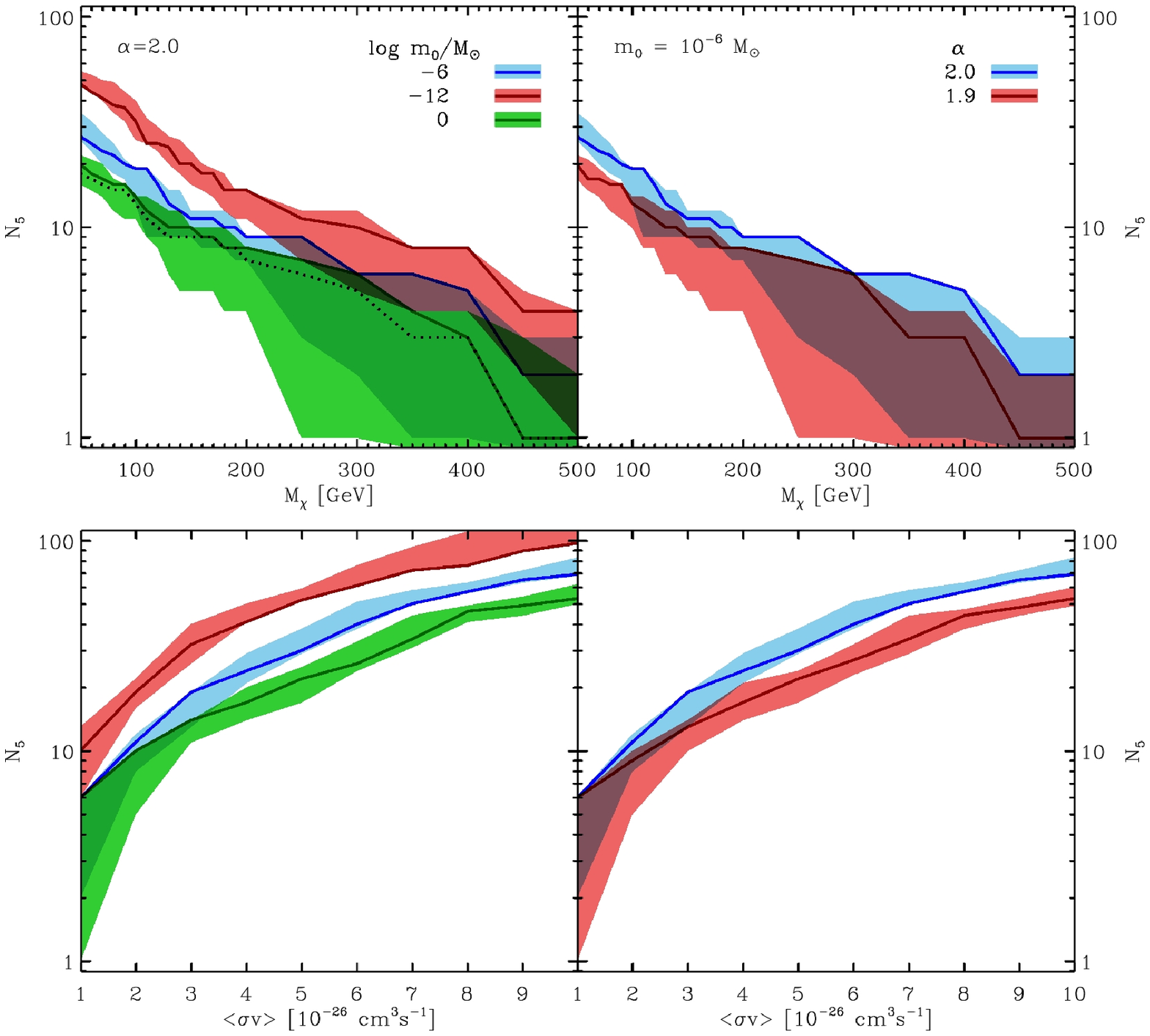}
\caption{$N_5$, the number of simulated subhalos exceeding $\sig=5$,
  as a function of the DM particle mass $\MX$ for $\sigv = 3 \times
  10^{-26}$ cm$^3$ s$^{-1}$ (\textit{top}) and the cross section
  $\sigv$ for $\MX=100$ GeV (\textit{bottom}). Dependence on the
  subhalo mass function cutoff mass $\mcut$ for slope $\alpha=2.0$
  (\textit{left}) and on $\alpha$ for $\mcut=10^{-6} \msun$
  (\textit{right}). The $\alpha=1.8$ case is almost identical to
  $\alpha=1.9$ and has been omitted from this plot. The shaded regions
  indicate the range of $N_5$ for ten randomly chosen observer
  locations and the solid lines refer to an observer placed along the
  intermediate axis of the host halo ellipsoid. The dotted line
  is the case without a boost factor. \label{fig:Npeaks}  }
\end{center}
\end{figure*}

\subsection{EGRET Constraints}\label{sec:egret_constraints}

Given the theoretically motivated modifications and additions to the
simulated flux distribution discussed in the previous sections, we
must ask whether our models still satisfy existing constraints from
EGRET on the diffuse gamma-ray flux as well as limits on a Galactic
center gamma-ray point source.

Integrating Eq.~\ref{eq:EGRET} from 30 MeV to 10 GeV, we obtain a
specific intensity of $5.91 \times 10^{-5}$ cm$^{-2}$ s$^{-1}$
sr$^{-1}$ for the extra-galactic background in the EGRET energy range.
Over this same energy range, DM annihilations result in 30-50 gamma
rays per event for $\MX = 50-500$ GeV(see
Section~\ref{sec:DM_parameters}). The diffuse flux plotted in
Fig.~\ref{fig:diffuse} can thus be converted to units of cm$^{-2}$
s$^{-1}$ sr$^{-1}$ by multiplying by $(3.09 \times 10^{21} {\rm
  cm/kpc}) \times (30-50) \times \sigv/\MX^2$. Even the strongest
diffuse background model considered here,
$(\alpha,\mcut/\msun)=(2.0,10^{-12})$, remains below the EGRET
extragalactic background for $\sigv/\MX^2 < 10^{-28}$ cm$^3$ s$^{-1}$
GeV$^{-2}$, i.e. over the full range of $\MX$ and $\sigv$ that we
consider in the following analysis.

\citet{Hooper2004} performed a re-analysis of the EGRET Galactic
center data with an improved energy dependent point spread function
and found no evidence of a point source at the location of Sag A$^*$.
For $\MX$ between 50 and 500 GeV they determined a 95\% confidence
upper limit of $\sim 10^{-8}$ cm$^{-2}$ s$^{-1}$ for gamma rays above
1 GeV.  The total flux from the innermost 30 arcmin (EGRET's angular
resolution) from a central cusp matched to VL-II's density profile at
$r_{\rm conv}$ is equal to $5.7 \times 10^{-2}$ GeV$^2$ kpc cm$^{-6}$
for a slope of $-1.0$. In the energy range from 1 to 30 GeV, DM
annihilations produce $8-33$ gamma rays per event for $\MX=50-500$
GeV, and the central flux exceeds the EGRET limit if $\sigv > 4.1
\times 10^{-26} (\MX/100{\rm\;GeV})^{1.41}$ cm$^{3}$ s$^{-1}$. Most of
the parameter space we considered here satisfies this limit, but
models with $\MX < 80$ GeV and $\sigv=3 \times 10^{-26}$ cm$^{3}$
s$^{-1}$ violate it. For an inner slope of $-1.2$ the constrait is
stronger, and the EGRET limit is already exceeded if $\sigv > 1.5
\times 10^{-26} (\MX/100{\rm\;GeV})^{1.41}$ cm$^{3}$ s$^{-1}$. In this
case a DM particle mass greater than 170 GeV (for $\sigv = 3 \times
10^{-26}$ cm$^3$ s$^{-1}$) would be required in order to remain below
the EGRET point source limit. For the uncorrected central flux, which
may in fact be closer to reality if dynamical processes have removed a
significant amount of DM from the Galactic center, the limit is
relaxed to $\sigv > 1.5 \times 10^{-25} (\MX/100{\rm\;GeV})^{1.41}$
cm$^{3}$ s$^{-1}$, in which case all models considered in this work
would satisfy the limit.

\begin{figure}
\begin{center}
\includegraphics[width=\columnwidth]{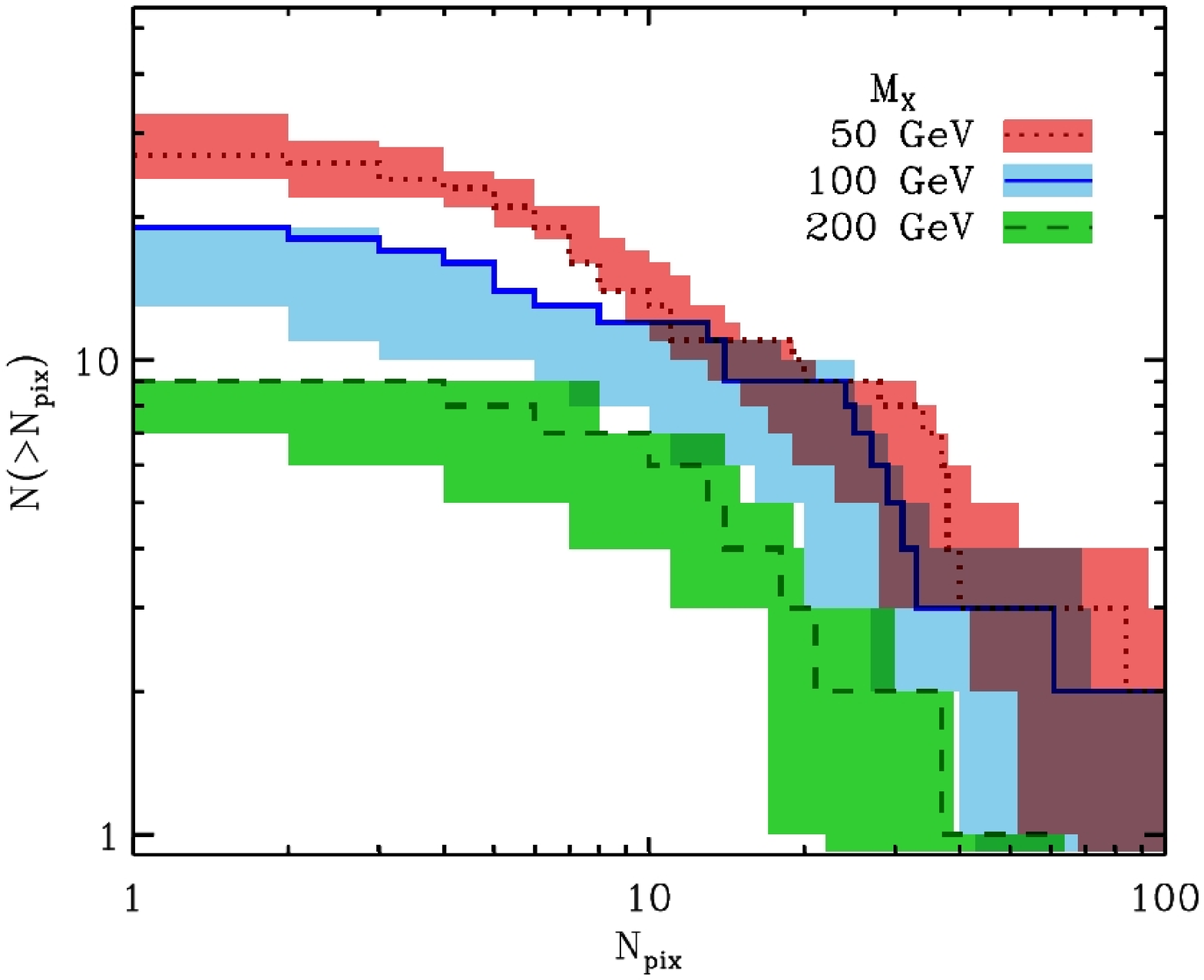}
\caption{The number of detectable ($\sig=5$) subhalos with more
  than $N_{\rm pix}$ detectable pixels versus $N_{\rm pix}$, for three
  different choices of $\MX$ (assuming $\sigv = 3 \times 10^{-26}$
  cm$^3$ s$^{-1}$). The shaded regions show the range of
  $N(>\!\!N_{\rm pix})$ for ten randomly chosen observer locations and
  the solid lines refer to an observer placed along the intermediate
  axis of the host halo ellipsoid.
  \label{fig:peak_sizes}}
\end{center}
\end{figure}

\begin{figure}
\begin{center}
\includegraphics[width=\columnwidth]{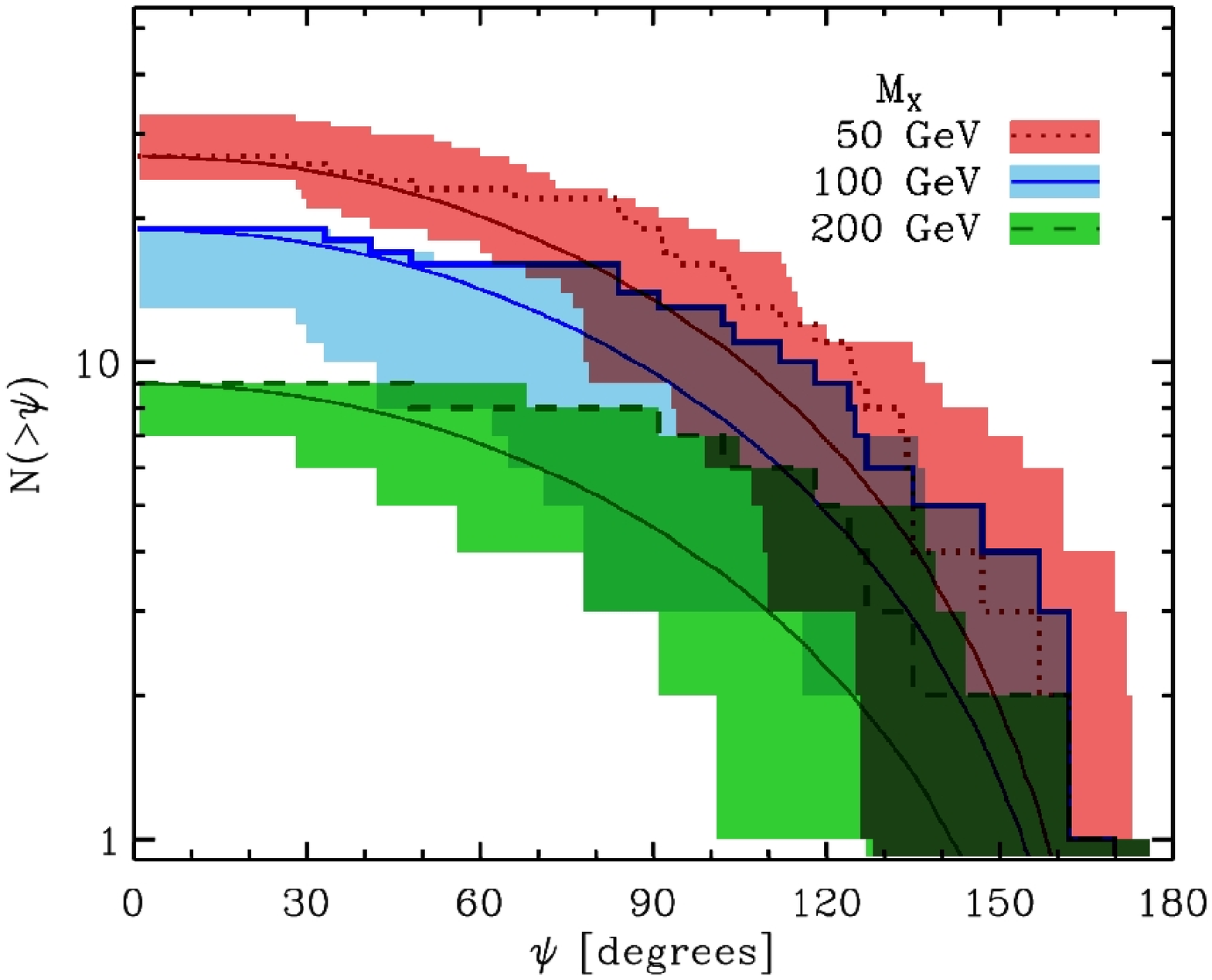}
\caption{The cumulative angular distribution of detectable ($\sig>5$)
  subhalos, for three different choices of $\MX$ (assuming $\sigv = 3
  \times 10^{-26}$ cm$^3$ s$^{-1}$). $\psi$ is the angle from the
  direction of the host center to the brightest pixel in each subhalo.
  The shaded regions show the range of $N(>\psi)$ for ten randomly
  chosen observer locations and the solid lines refer to an observer
  placed along the intermediate axis of the host halo ellipsoid. The
  thin solid line represents an isotropic distribution normalized to
  the fiducial observer's location.
  \label{fig:peak_angles}}
\end{center}
\end{figure}

\subsection{Detectable subhalos}\label{sec:detectable}

We have now assembled all necessary ingredients to convert our
simulated particle distribution into a quantitative prediction for the
number of subhalos detectable with GLAST. To recap: $N_s$, the
expected number of gamma-rays from a subhalo in one GLAST 9 arcmin
``pixel'', is calculated by summing $B(M) \rho_i m_p / (4\pi d_i^2)$
over all the subhalo's particles falling within the pixel and
multiplying by $\tau_{\rm exp} \sigv/(2\MX^2) \int_{E_{\rm th}}^{\MX}
dN_\gamma/dE \, A_{\rm eff}(E) \, dE$. The total signal from a subhalo
is obtained by summing $N_s$ over all $S/N>1$ pixels covered by the
subhalo. This flux is then compared to $\sqrt{\sum N_b}$, the square
root of the expected number of background photons, made up of an
isotropic extragalactic astrophysical component (as measured by
EGRET), an astrophysical Galactic background (calculated with
GALPROP), and two diffuse DM annihilation backgrounds, one from
undetectable subhalos and one from the smooth host halo. The resulting
detection significance $\sig=\sum N_s/\sqrt{\sum N_b}$ depends
critically on the mass of the DM particle (smaller $\MX$ is better)
and on its annihilation cross section (larger $\sigv$ is better). At
fixed $\MX$ and $\sigv$, $\sig$ depends on the normalization, the
slope $\alpha$, and the low mass cut-off $\mcut$ of the subhalo mass
function through the calculation of the boost factor and the two
diffuse annihilation backgrounds. Note that changes in $\alpha$ and
$\mcut$ affect $\sig$ in opposite ways: increasing the abundance of
subhalos below VL-II's resolution limit, by increasing $\alpha$ or
lowering $\mcut$, raises the boost factor (good for detection) as well
as the background from undetectable subhalos (bad for detection).

Due to departures from spherical symmetry in the host halo mass and
subhalo spatial distribution \citep{Kuhlen2007}, the number of
detectable subhalos will depend on the observer position. For an
observer located at 8 kpc along the major axis of the host halo
ellipsoid the diffuse host halo background will be higher than for an
observer at the same distance along the minor axis
\citep{CalcaneoRoldan2000}. Individual bright subhalos may also become
undetectable from certain vantage points, if they fall behind the
Galactic disk or center for example. For these reasons we have
performed our analysis for ten randomly drawn observer positions, in
addition to our fiducial observer located along the host halo
intermediate axis. Our results are presented in Fig.~\ref{fig:Npeaks},
where we plot $N_5$, the number of VL-II subhalos exceeding $\sig=5$,
as a function of $\MX$ and $\sigv_{-26} = \sigv/(10^{-26}$ cm$^3$
s$^{-1}$), for different choices of $\alpha$ and $\mcut$.

We find that $N_5$ ranges from a few up to multiple tens over the
range of $\MX$ and $\sigv$ we have considered. In our fiducial model,
$(\MX/{\rm GeV},\sigv_{-26},\alpha,\mcut)=(100,3,2.0,10^{-6}$), an
observer positioned along the host halo's intermediate axis would be
able to detect $N_5=19$ subhalos. For the set of ten randomly placed
observers we find a range of $N_5$ from 13 to 19. Lowering $\mcut$
from $10^{-6} \msun$ to $10^{-12} \msun$, the smallest value found by
\citet{Profumo2006}, results in a factor $\sim 2$ increase in $N_5$.
In this case, assuming $\alpha=2.0$, more than ten subhalos should be
detectable even with $\MX$ as high as 300 GeV (for $\sigv_{-26}=3)$ or
$\sigv$ as low as $10^{-26}$ cm$^3$ s$^{-1}$ (for $\MX=100$ GeV). If,
on the other hand, no subhalos exist with masses less than $1 \msun$,
then $N_5$ drops below 10 at $\MX \approx 160$ GeV (for
$\sigv_{-26}=3)$ or $\sigv_{-26} \approx 2$ (for $\MX=100$ GeV).
Reducing $\alpha$ below the critical value of 2.0 suppresses $N_5$ by
about a factor of 2/3. At a fiducial value of $(\MX,\sigv_{-26}) =
(100,3)$ we find $N_5=13-19$, $26-40$, $11-14$, $10-14$, and $9-14$
for $(\alpha,\mcut)=(2.0,10^{-6})$, $(2.0,10^{-12})$, $(2.0, 10^0)$,
$(1.9, 10^{-6})$, and $(1.8, 10^{-6})$, respectively.  In the most
favorable case considered here, $(\MX/{\rm
  GeV},\sigv_{-26},\alpha,\mcut)=(100,10,2.0,10^{-12}$), the number of
detectable subhalos exceeds 100. The dependence of $N_5$ on $\mcut$
and $\alpha$ is primarily driven by their effect on the boost factor:
lowering the abundance of DM subhalos below VL-II's resolution limit
reduces the boost factor and lowers $N_5$. For comparison we have also
plotted results without a boost factor for the
$(\alpha,\mcut)=(2.0,10^{-6}$) case. This case is similar to the
$\mcut=1 \msun$ case, and even without invoking such a
sub-substructure boost factor multiple subhalos should be bright
enough for detection for most of the parameter space probed here.

The set of models we have considered includes a model similar to the
one recently proposed by \citet{Hooper2007} to explain the excess
microwave emission around the Galactic center measured by WMAP
\citep[so-called ``WMAP haze'',][]{Finkbeiner2004}. \citet{Hooper2007}
showed that the intensity and angular distribution of the WMAP haze
can be modeled as synchrotron emission from relativistic electrons and
positrons generated in dark matter annihilations, assuming a cusped
central DM density profile of slope $-1.2$, a DM particle mass of
$\sim 100$ GeV, and an annihilation cross section of $\sigv = 3 \times
10^{-26}$ cm$^3$ s$^{-1}$. If this explanation for the WMAP haze turns
out to be correct, we predict that GLAST should be able to detect
about 15 subhalos.

In Fig.~\ref{fig:peak_sizes} we present the cumulative size
distributions of all detectable subhalos, for $\MX=50$, 100, and 200
GeV at fixed $\sigv = 3 \times 10^{-26}$ cm$^3$ s$^{-1}$. The size
here is defined as the number of $S/N>1$ pixels ($N_{\rm pix}$)
contributing to the total subhalo signal exceeding the detection
threshold of $\sig=5$. Recall that we have matched the pixel size to
GLAST's angular resolution of 9 arcmin, such that each pixel
corresponds to a solid angle of $4.363 \times 10^{-6}$sr. Again we
plot distributions for the fiducial observer on the host halo's
intermediate axis as well as the minimum and maximum over all ten
randomly drawn observer positions. Detectable subhalos are typically
extended, with a median of $N_{\rm pix}$ of 13. Only about 5\% are
detected in only one pixel. This should aid in discriminating between
subhalos lit up by DM annihilation and conventional astrophysical
sources (e.g. gamma-ray pulsars) which will often appear as point
sources \citep{Baltz2007}.

The angular distribution of detectable subhalos is shown in
Fig.~\ref{fig:peak_angles}, where we plot the cumulative number of
detectable subhalos more than $\psi$ degrees from the direction
towards the host center. For the fiducial observer we have overplotted
an isotropic distribution (thin solid line), in which detectable
subhalos are uniformly distributed over the whole sky. The actual
distribution indicates a slight excess at large angles compared to an
isotropic distribution. However, a KS test shows that this discrepancy
is not statistically significant. For only two of all eleven observer
positions can the null hypothesis of an isotropic distribution be
rejected at more than 90\% confidence, and we conclude that the
distribution of detectable sources on the sky is consistent with
isotropy.

\begin{figure}
\begin{center}
\includegraphics[width=\columnwidth]{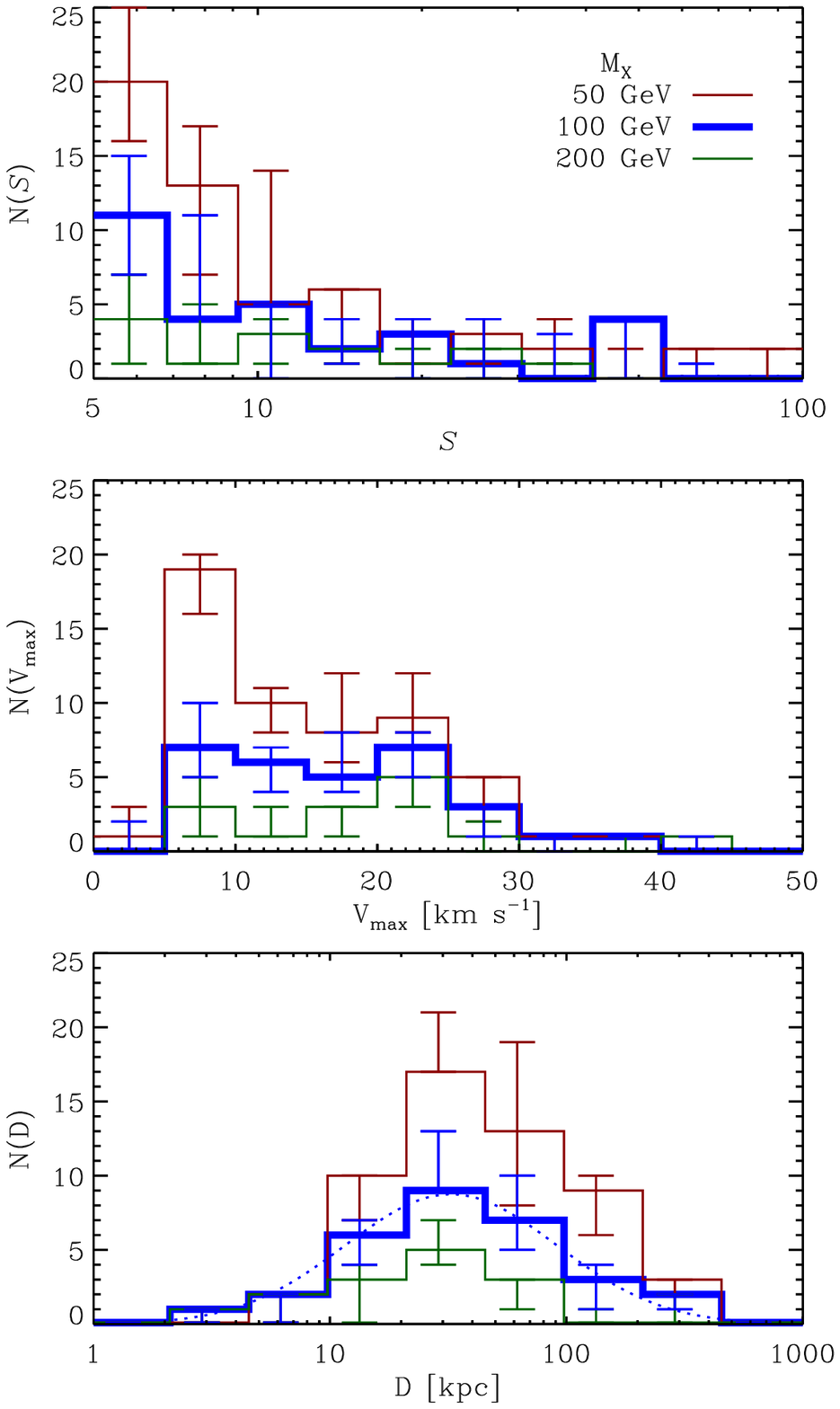}
\caption{Differential distribution of detection significance $\sig$
(\textit{top}), $\Vmax$ (\textit{center}), and distance to the observer
(\textit{bottom}) of the subset of detectable ($\sig>5 $) subhalos,
for three different choices of $\MX$ (assuming $\sigv = 5 \times
  10^{-26}$ cm$^3$ s$^{-1}$). The solid lines are for the fiducial
  observer placed along the intermediate axis of the host halo
  ellipsoid and the error bars indicate the range of values for 10
  randomly chosen observer locations. For the distance distribution the best-fit Gaussian has been overplotted for the $\MX=100$ GeV case.
  \label{fig:peak_distributions}}
\end{center}
\end{figure}

Next we consider the differential distribution of the detection
significance $\sig$, $\Vmax$, and the distance to the observer, of the
subset of detectable subhalos, see Fig.~\ref{fig:peak_distributions}.
In order to obtain better statistics for this analysis we chose a
slightly higher cross section of $\sigv=5 \times 10^{-26}$ cm$^3$
s$^{-1}$, resulting in 51, 29, and 12 detectable subhalos for $\MX=$
50, 100, 200 GeV, respectively. The histograms represent the results
for the fiducial observer position and the error bars indicate the
range of values for the 10 randomly drawn positions.

The $\sig$ distribution is slightly peaked towards lower $\sig$
values, with about two thirds of all detectable subhalos having
$\sig<10$. About 10\% of subhalos should be detectable at very high
significances, with $\sig$ exceeding 25. For these, the median $\Vmax$
is 24 km s$^{-1}$, indicating that the most detectable subhalos tend
to be the more massive ones. These would be more likely to host a
dwarf galaxy \citep[e.g.][]{Madau2008}, which in some cases may have
so far eluded discovery due to their ultra-faint optical surface
brightness. The $\Vmax$ distribution, however, reveals that a
siginificant number of lower $\Vmax$ subhalos should also be
detectable. From 5 to 25 $\kms$ the $\Vmax$ distribution is
approximately flat, and in the $\MX=50$ GeV case it even exhibits a
pronounced peak towards lower $\Vmax$. No detectable subhalos are
found with $\Vmax$ less than 5 km s$^{-1}$, which is probably due to a
lack of numerical resolution in these low mass halos. It appears
likely that the total number of detectable subhalos has not yet
converged in our simulations. We explore this further in
Section~\ref{sec:convergence} below.

The distance distribution of detectable subhalos is broadly peaked
around the median of 40 kpc from the observer. For the $\MX=100$ GeV
case, it is consistent with a simple log-normal distribution, centered
at 32 kpc with a width of $\sigma_{\log_{10}\!D}=0.45$. About 80\% of
all detectable subhalos have distances between 10 and 100 kpc, and
only 7.5\% are closer than 10 kpc.


\subsection{Convergence?}\label{sec:convergence}

In Section~\ref{sec:backgrounds} we discussed the diffuse background
resulting from undetectable subhalos, but we should also consider
whether any subhalos below VL-II's resolution limit of $10^6 \msun$
might be detectable as individual extended sources. The fact that the
$\Vmax$ distribution of detectable subhalos remains flat (or even
rises slightly) down to the smallest visible subhalos hints that the
number of detectable subhalos has not yet converged in our
simulations. Further evidence comes from directly comparing the VL-II
results with the lower resolution VL-I simulation. In VL-I we find far
fewer detectable subhalos. In fact, with $\sigv=3 \times 10^{-26}$
cm$^3$ s$^{-1}$, at most 15 subhalos are visible, and that only for
the most optimistic assumptions. Typically only VL-I's most massive
subhalo reaches $\sig=5$. The likely explanation for the dearth of
detectable subhalos in VL-I is that the lower numerical resolution
resulted in a significant suppression of substructure close to the
center. In VL-I only 42 subhalos with at least 200 particles are
closer than 50 kpc from the halo center, whereas in VL-II this number
is 362, of which 108 are more massive than $4.2 \times 10^6 \msun$
(200 VL-I particles). Subhalos in VL-II also typically have higher
central densities: the median of the maximum density in all subhalos
with $\msub>4.2 \times 10^6 \msun$ is 5.4 times higher in VL-II than in
VL-I.  This will lead to higher fluxes and hence increased
detectability.

These considerations motivate a calculation of the total expected
number of detectable subhalos below the simulation's resolution limit,
including the full hierarchy of substructure all the way down to the
free streaming cutoff $\mcut$.  Microhalos, the remnants of the
earliest collapsed DM structures, are of particular interest here, as
they have recently been suggested as possible GLAST sources that may
even exhibit proper motion \citep{Koushiappas2006}. In the following
discussion we focus on resolved subhalos, because above the VL-II
resolution the unresolved component is small, and, as we now show, the
fraction of unresolved, detectable sources will further decrease with
mass.

In the previous section we showed that in our simulations 95\% of all
detectable subhalos are extended sources covering more than one pixel.
The fraction of point-like source is likely to become even smaller for
less massive systems. For an NFW density profile, 87.5\% of the
annihilation luminosity is generated within the scale radius, hence
$r_s/D$ is a good proxy for a subhalo's angular size. Assuming an
angular resolution of 9 arcmin, the maximum distance out to which a
subhalo will appear extended is $D_{\rm max} = r_s / 9$ arcmin $= 380
\; r_s$.  For a \citet{Bullock2001} concentration-mass relation $c(M)
\propto M^{-0.025}$ for halos with masses below $10^6 \msun$. In this
case
\begin{equation}
D_{\rm max} \propto r_s \propto R/c \propto M^{1/3} M^{0.025} \propto
M^{0.358}.
\end{equation}
Specifically, assuming that local subhalos are three times more
concentrated than field halos (Eq.~\ref{eq:c_of_r}), we find
\begin{equation}
D_{\rm max} \approx 0.6 \left( \f{M}{10^{-6} \msun} \right)^{0.358} \!\!
{\rm pc}.
\label{eq:Dmax}
\end{equation}
At distances exceeding $D_{\rm max}$, sources appear point-like and
the surface brightness in one pixel simply scales as $L/D^2$. For
microhalos with $M<1 \msun$, $L \propto M^{0.93}$ and hence the
maximum surface brightness of such unresolved sources scales with mass
as $L/D_{\rm max}^2 \propto M^{0.93} \times (M^{0.358})^{-2} \propto
M^{0.214}$, resulting in a smaller fraction of detectable point-like
sources for lower mass subhalos.

\begin{figure}
\begin{center}
\includegraphics[width=\columnwidth]{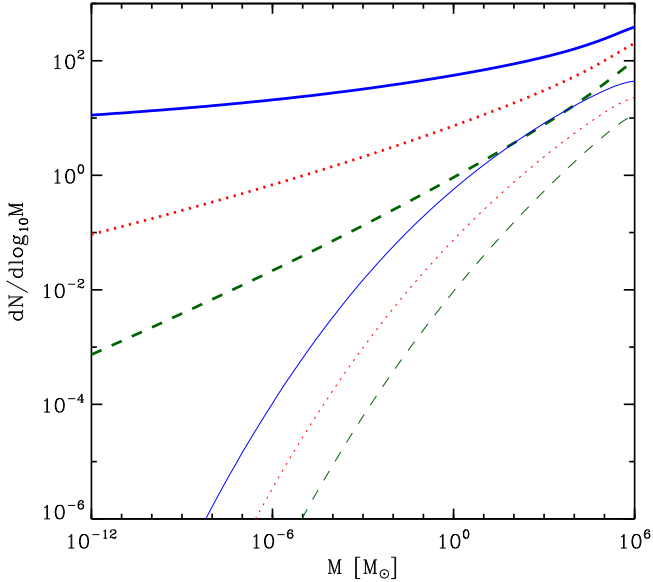}

\caption{The differential mass function of extended (i.e.
  GLAST-resolvable, $r_s/D>9$ arcmin, thick upper lines) and
  observable ($r_s/D>9$ arcmin \textit{and} $\Phi>1$ GeV$^2$ kpc
  cm$^{-6}$ sr$^{-1}$, thin lower lines) subhalos below VL-II's mass
  resolution of $10^6 \msun$, for $\alpha=2.0$ (solid), 1.9 (dotted),
  and 1.8 (dashed).
  \label{fig:microhalos}}
\end{center}
\end{figure}

The differential mass function of extended (i.e. GLAST-resolvable)
subhalos scales approximately as $dN_{\rm ext}/dM \propto dn_{\rm
  sub}/dM D_{\rm max}^3 \sim M^{-\alpha+1.074}$, implying that even
for $\alpha=2.0$ the total number $N_{\rm ext}$ will be dominated by
the most massive subhalos, i.e.  those just below VL-II's resolution
limit. The results of a more realistic calculation (details in
Appendix~\ref{sec:AppendixB}), taking into account the radial
concentration bias and a log-normal concentration distribution at a
given mass, are consistent with this simple scaling and are shown with
the thick upper lines in Fig.~\ref{fig:microhalos}.  Integrating this
differential extended source mass function from $\mcut=10^{-6}\msun$
to $10^6\msun$, we find $N_{\rm ext}=1090, 324, 115$ for $\alpha =
$2.0, 1.9, and 1.8, respectively.

Of course not all of these extended subhalos would actually be
detectable by GLAST. We can extend the above calculation by imposing
an additional constraint on the subhalo surface brightness. For an NFW
density profile the surface brightness within $r_s$ is given by
\begin{equation}
\Phi = \f{L}{4 \pi D^2 \pi (r_s/D)^2} = \f{L}{4 \pi^2 r_s^2} \approx \f{0.875}{3\pi} \rho_s^2 r_s.
\end{equation}
Since $\rho_s \propto c^3/f(c)$, $\Phi \propto R c^5/f(c)^2 \propto
M^{1/3} c^5/f(c)^2$.  Thus for a halo of a fixed mass, the surface
brightness has a very steep concentration dependence, $\Phi \propto
c^{4.316}$ for $M<10^6 \msun$.  For this reason it is very important
to account for the radial bias and the log-normal scatter in subhalo
concentrations. A conservative choice for the limiting surface
brightness for detection is $\Phi_0=1$ GeV$^2$ kpc cm$^{-6}$
sr$^{-1}$, which is a bit below the minimum central brightness of all
detectable subhalos in our study.  The thin lower lines in
Fig.~\ref{fig:microhalos} show how this additional constraint lowers
the expected number of detectable extended microhalos. Lower
mass halos are more strongly affected by a surface brightness cut,
since $\Phi$ scales as $M^{0.225}$ for $c \propto M^{-0.025}$. In
total we find the surface brightness constraint lowers the expected
number of detectable subhalos with masses below $10^6\msun$ by about
an order of magnitude to 82, 32, and 13 subhalos for $\alpha=$2.0,
1.9, and 1.8, respectively. Thus we conclude that subhalos below
VL-II's resolution limit contribute at most 3/4 of the expected total
number of detectable extended sources.

Contrary to previous results \citep{Koushiappas2006}, we find that
GLAST is unlikely to be able to detect and resolve individual
microhalos, defined here as subhalos with masses less than $1 \msun$.
According to our calculations, only 0.40, 0.042, and 0.0044 subhalos
with masses between $10^{-6}$ and $1 \msun$ would satisfy both angular
size and surface brightness constraints. The disagreement between
\citet{Koushiappas2006} and our results is explained by our lower
local subhalo abundance ($\xi = (M^2 dn/dM) / \rho_{\rm host}(R_\odot)
= 8\times10^{-4}$ versus his 0.002), higher local concentrations
resulting in smaller angular sizes, and a more realistic treatment of
the expected backgrounds, including a contribution from undetectable
subhalos. For comparison, we find that the local abundance of
microhalos is (10.7, 0.388, 0.0137) pc$^{-3}$ for $\alpha=(2.0, 1.9,
1.8)$ for the anti-biased radial subhalo distribution and (189, 6.83,
0.241) pc$^{-3}$ in the unbiased case.

\section{Discussion and Conclusion}\label{sec:conclusion}

In this work we have considered the possibility of directly observing
Galactic DM substructure through the detection of gamma rays
originating in the annihilation of DM particles in the centers of
subhalos. Based on a hybrid approach, making use of both the highest
resolution numerical simulations available as well as analytical
models for the extrapolation beyond this simulation's resolution
limit, we have shown that for reasonable values of the DM particle
physics and subhalo mass function parameters, future gamma-ray
observatories, such as the soon to be launched GLAST satellite, may
very well be able to detect a few, even up to several dozen, such
subhalos.

An important overall systematic uncertainty in our study is the nature
of the DM particle. Our results are valid for the case of a weakly
interacting relic particle, such as the lightest supersymmetric
particle in supersymmetric extensions of the standard model. If the DM
particle instead turns out to be an axion, for example, then DM
annihilation would not occur and our results would be irrelevant. Even
in the WIMP case, the allowed parameter space for the mass and
annihilation cross section of the DM particle spans many orders of
magnitude. Individual subhalos will only be detectable with GLAST, if
$\MX$ and $\sigv$ fall in an ``observable window'', roughly given by
$\MX \sim 50-500$ GeV for $\sigv \sim 1-10 \times 10^{-26}$ cm$^3$
s$^{-1}$.

Another set of uncertainties is associated with our use of a numerical
simulation. First of all, we have only simulated one host halo at very
high resolution. Statistical studies at lower resolution
\citep[e.g.][]{Reed2005} have found about a factor of two halo-to-halo
scatter in the substructure abundance. Secondly, an uncertainty arises
from the importance of DM substructure below the resolution limit of
our simulation, due to its boosting effect on the brightness of
individual subhalos, its contribution to a diffuse background from
undetectable subhalos, and as individually detectable sources. Lastly,
the lack of a baryonic component in our simulation is another source
of uncertainty. In addition to the effects baryonic physics might have
on the DM distribution at the Galactic center (see
Section~\ref{sec:central_flux_corrections}), one may worry that the
abundance or detectability of nearby subhalos could be reduced by
passages through the Milky Way's stellar disk. We can use the subhalo
orbits from our simulations to constrain how large an effect such disk
crossings would have.

For this estimate we have used orbits from the VL-I simulation
\citep{Diemand2007b,Kuhlen2007}, as we are still in the process of
extracting them from the VL-II dataset. We model the Milky Way's
stellar disk as an exponential disk with scale radius $R_d=3.5$ kpc
and scale height $z_d=350$ pc. Only 8.2\% of all VL-I subhalos ever
entered a central disk region delimited by four scale lengths, i.e. $R
< 4 R_d$ and $|z| < 4 z_d$.\footnote{For simplicity we aligned the
  disk with the z-axis, but the results are insensitive to this
  choice.}  This fraction grows to 22\% when only subhalos within 50
kpc of the halo center at $z=0$ are considered. Only about one fifth
of all disk crossing subhalos do so more than once. The median of
$V_z$, the velocity component perpendicular to the disk, is quite
large, about $400 \kms$.  Since the typical disk crossing times are
thus a few Myr, much less than the internal dynamical time of subhalos
even within $r_s$ ($\sim 100$ Myr), we can apply the impulse
approximation to determine how much subhalos will be heated by disk
passages \citep[see \S 7.2 of][]{Binney1987}. The change in the
z-component of the velocity of a DM particle belonging to a disk
crossing subhalo is given by
\begin{equation}
  \Delta V_z = \f{2 |z|}{V_z} |g_z(R)|,
\end{equation}
where $|z|$ is the height of the particle above the subhalo's midplane
and $g_z(R)$ is the gravitational field of the disk at radius R, and
is approximately equal to $g_z(R) = 2 \pi G \Sigma_0
\exp[(R_0-R)/R_d]$.  Here $\Sigma_0 \simeq 75 \msun$ pc$^{-2}$ is the
surface density of the disk in the solar neighborhood $R=R_0$. Setting
$z=\rVmax$, we find
\begin{eqnarray}
  \Delta V_z(|z|) & = & 5 \; \kms \exp\left[\f{R_0-R}{R_d}\right] \nonumber \\
 & & \times \left(\f{|z|}{500 {\rm pc}}\right) \left(\f{V_z}{400 \kms}\right)^{-1}.
\end{eqnarray}
We can understand how destructive such kicks will be to the subhalo by
comparing $\Delta V_z$ to the local circular velocity at height $z$.
We find that for all disk crossings in VL-I the median of $(\Delta
V_z/V_c)$ is 0.4 at $z=r_s$, and only 25\% have $\Delta V_z > V_c$ at
$r_s$.  These numbers will be even smaller closer to the subhalo's
center, since $\Delta V_z \propto z$ and $V_c \propto \sqrt{r}$. We
conclude that only about 5\% of all subhalos within 50 kpc today might
have experienced a significant reduction in central density, and hence
in annihilation luminosity, from disk crossings.

Given the substantial uncertainties discussed above, one may wonder
whether it is even sensible to consider DM annihilations from subhalos
at all. We believe that the mere possibility of detecting DM
annihilations with GLAST offers such an exciting opportunity to
directly confirm the existence of a DM particle and to learn something
about its properties, that it warrants theoretical investigations such
as the present one.

We conclude by summarizing our main findings:

\begin{itemize}

\item 
  Numerical simulations indicate that DM is far from smoothly
  distributed throughout the Galactic halo. Extremely high resolution
  simulations, such as the Via Lactea series, have shown that this
  clumpiness extends into individual subhalos, resulting in
  sub-substructure. Since the annihilation rate is proportional to the
  DM density squared, such clumpiness leads to a boost by $(1+B(M))$
  of the annihilation luminosity compared to a smooth density
  distribution. Any determination of the gamma-ray brightness of
  individual subhalos must account for this boost factor. In the case
  of simulated subhalos only the portion of the sub-substructure
  hierarchy below the simulation's resolution limit should be included
  in the boost factor calculation. Our analytical model of the boost
  factor, which assumes a powerlaw subhalo mass function of slope
  $\alpha$ and low-mass cutoff $\mcut$, a resolved subhalo mass
  fraction of 10\%, and a \citet{Bullock2001} like concentration-mass
  relation, results in moderate boost factors of a few up to 10 for
  $(\alpha,\mcut)=(2.0,10^{-6})$. Lowering $\mcut$ to $10^{-12}$
  raises $B(M)$ by a factor of 2. For $\alpha=1.9$ $B(M)$ barely
  exceeds unity.

\item 
  The portion of the substructure hierarchy below Via Lactea's
  resolution also contributes to a diffuse annihilation background
  from undetectable sources. The magnitude of this background
  depends on the spatial distribution of subhalos within the host halo
  in addition to the subhalo mass function parameters
  $(\alpha,\mcut)$. We have extended the model developed by
  \citet{Pieri2008} to allow for an anti-biased subhalo radial
  distribution and for higher subhalo concentrations closer to the
  host halo center. The resulting background tends to dominate the
  diffuse background from the smooth host halo at angles greater than
  $\psi=10^\circ-60^\circ$, depending on the choice of
  $(\alpha,\mcut)$. At high galactic latitudes this background can
  dominate the astrophysical backgrounds (both Galactic and
  extragalactic) and should be accounted for when determining subhalo
  detectability.

\item 
  By comparing the expected number of gamma-ray photons from an
  individual subhalo to the square root of the number of expected
  background photons, we can determine a detection significance $\sig$
  for each subhalo. $\sig$ depends strongly on the mass $\MX$ and
  annihilation cross section $\sigv$ of the DM particle as well as on
  the subhalo mass function parameters $(\alpha,\mcut)$.  We find that
  for $\MX \sim 50-500$ GeV and $\sigv \sim 1-10 \times 10^{-26}$
  cm$^3$ s$^{-1}$, a few subhalos, even up to a few tens, exceed
  $\sig=5$, and hence should be detectable with GLAST. In the most
  optimistic case we considered ($\MX=100$ GeV, $\sigv=10^{-25}$
  cm$^3$ s$^{-1}$, $\alpha=2.0$, and $\mcut=10^{-12} \msun$) almost
  100 subhalos would be detectable, whereas increasing $\MX$ to 500
  GeV or lowering $\sigv$ to $10^{-26}$ cm$^3$ s$^{-1}$ sharply
  reduces the expected number of sources, especially for low $\alpha$
  or high $\mcut$.

\item 
  For the particular DM annihilation model that \citet{Hooper2007}
  proposed to explain the WMAP haze, an excess microwave emission
  around the Galactic center, namely $\MX \sim 100$ GeV, $\sigv=3
  \times 10^{-26}$ cm$^3$ s$^{-1}$, and a central host halo density
  cusp of slope -1.2, we find that GLAST should be able to detect
  around 15 subhalos, for reasonable choices of $(\alpha,\mcut)$.

\item 
  Th majority of all detectable subhalos are resolved with GLAST's
  expected angular resolution of 9 arcmin. The median number of pixels
  is 13, and only 5\% are detected in only one pixel.

\item 
  For 9 out of the 11 observer locations we considered, the angular
  distribution of detectable subhalos is consistent with an isotropic
  distribution on the sky.

\item 
  Those subhalos that are detected with the highest significances
  ($\sig>25$) tend to be the more massive ones, with a median $\Vmax$
  of 24 km s$^{-1}$. However, the $\Vmax$ distribution of detectable
  subhalos is flat, or slightly rising, towards lower $\Vmax$, down to
  $\Vmax=$ 5 km s$^{-1}$.

\item 
  Detectable subhalos typically lie between 10 and 100 kpc from the
  observer. The distance distribution is consistent with a log-normal
  distribution centered at 32 kpc and a width of
  $\sigma_{\log_{10}\!D}=0.45$.

\item It appears that the number of detectable subhalos has not yet
  converged in our simulations. Higher resolution simulations would
  likely resolve smaller clumps, which are more abundant and more
  likely to be found near the observer. We estimate that we may be
  missing up to three quarters of all detectable subhalos.

\end{itemize}

Our results suggest that searching for Galactic DM substructure should
be an important part of the GLAST data analysis in the upcoming years.

\acknowledgments 

We would like to thank our collaborators Ben Moore, Doug Potter,
Joachim Stadel, and Marcel Zemp for their help and assistance with
code development and testing, and for allowing the use of the Via
Lactea II dataset prior to publication. This work benefited from
fruitful discussions with Bill Atwood, Sergio Colafrancesco, Miguel
S\'anchez-Conde, Dan Hooper, Robert Johnson, Savvas Koushiappas,
Stefano Profumo, Louie Strigari, Andrew Strong, and Scott Tremaine.
Support for this work was provided by NASA grants NAG5-11513 and
NNG04GK85G. M.K.  gratefully acknowledges support from the Hansmann
Fellowship at the Institute for Advanced Study. J.D.  acknowledges
support from NASA through Hubble Fellowship grant HST-HF-01194.01. The
``Via Lactea I'' simulation was performed on NASA's {\it Project
  Columbia} supercomputer. ``Via Lactea II'' was performed on the {\it
  Jaguar} Cray XT3 supercomputer at the Oak Ridge National Laboratory,
through an award from DOE's Office of Science as part of the 2007
Innovative and Novel Computational Impact on Theory and Experiment
(INCITE) program. It is a pleasure to thank Bronson Messner and the
Scientific Computing Group at the National Center for Computational
Sciences for their help and assistance.

{}

\pagebreak

\begin{appendix}

\section{Calculation of the diffuse flux from individually undetectable subhalos}\label{sec:AppendixA}

\begin{figure}
\begin{center}
\includegraphics[width=0.6\textwidth]{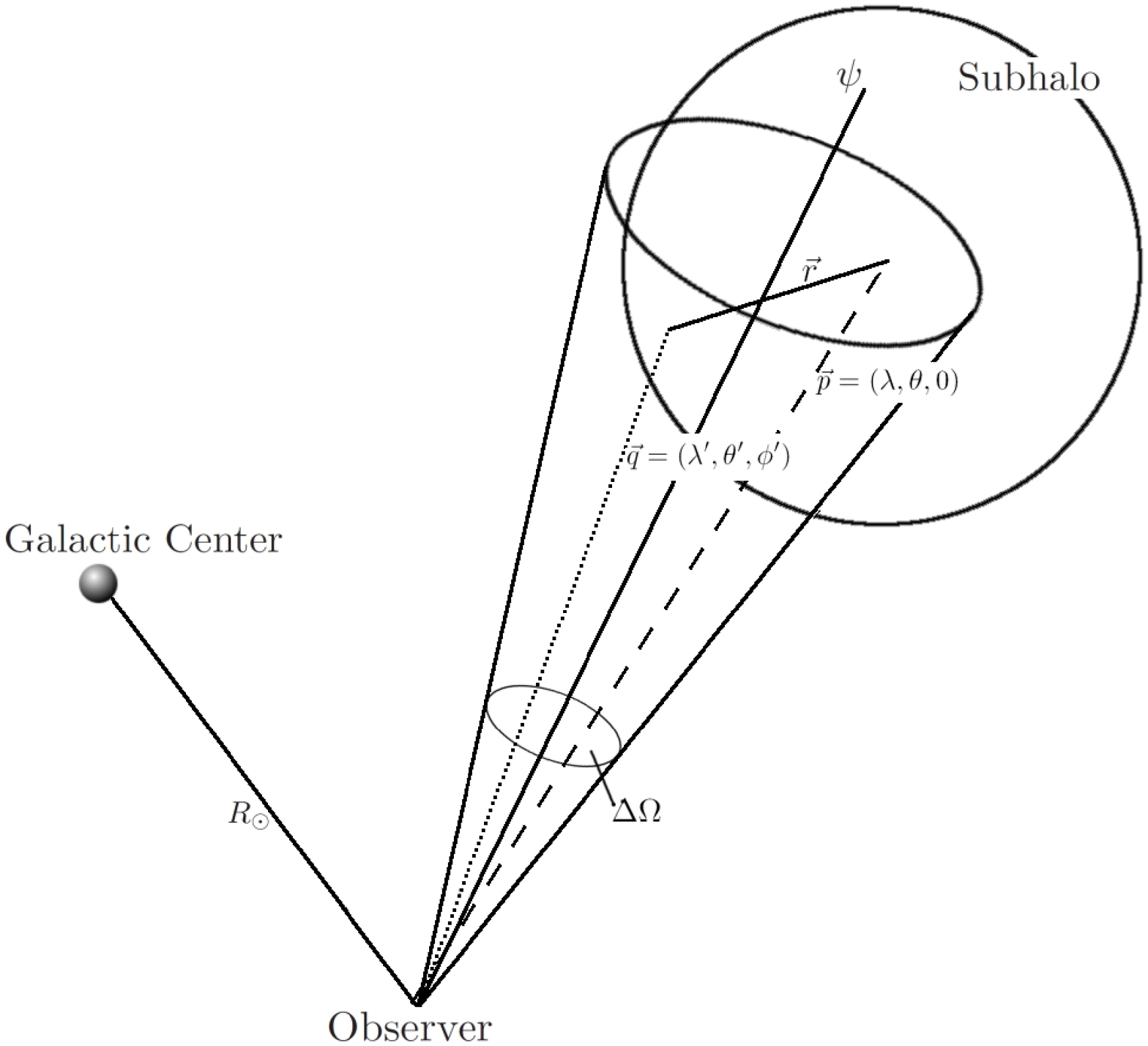}
\caption{Diagram of the relevant geometry for the calculation of $\mathcal{F}_{\rm halo}$ (Eq.~\ref{eq:phi_halo}).\label{fig:diagram} }
\end{center}
\end{figure}

We follow the methodology presented in \citet{Pieri2008}, and define
the diffuse flux from individually undetectable subhalos from a solid
angle $\Delta\Omega$ around the direction of observation $\psi$ by
\begin{equation}
\mathcal{F}(\psi,\Delta\Omega) = \int_{\Delta\Omega} d\!\cos\theta\;d\phi \int_M dM \int_{\rm l.o.s.} \hspace{-0.15in} d\lambda \; \lambda^2 \; \int_c dc \; P(c;M,R) \; \mathcal{F}_{\rm halo}(M,c,\lambda,\Delta\Omega) \; \f{dn_{\rm sub}}{dM}(M,R),
\label{eq:phi1}
\end{equation}
where $\lambda$ is the distance along the line of sight and
$R(\psi,\lambda,\theta,\phi)$ is the Galacto-centric distance, given by
\begin{equation}
R(\psi,\lambda,\theta,\phi) = \sqrt{\lambda^2 + R_\odot^2 -  2 \lambda R_\odot (\cos\theta\cos\psi - \cos\phi\sin\theta\sin\psi)}.
\end{equation}
The subhalo mass function and spatial distribtution is given by
\begin{equation}
\f{dn_{\rm sub}}{dM}(M,R) = A M^{-\alpha} \f{1}{(1+R/r_s^{\rm MW})^2} \msun^{-1} \; {\rm kpc}^{-3},
\end{equation}
and we have normalized it such that the total mass in subhalos with
masses between $m_0=10^{-5} \mvir$ and $m_1=10^{-2} \mvir$ is 10\% of
$\mvir$, i.e.
\begin{equation}
\int_{m_0}^{m_1} dM M \int_0^{\rtwo} \!\! 4\pi R^2 \f{dn_{\rm sub}}{dM}(M,R) \; dR = 0.1 \mvir.
\end{equation}

Note that this anti-biased spatial distribution is motivated by
numerical simulations \citep{Kuhlen2007, Madau2008} and differs from
the distribution chosen by \citet{Pieri2008}, who employed an unbiased
distribution of the form
\begin{equation}
\f{dn_{\rm sub}}{dM}(M,R) = A M^{-2} \f{\mathcal{H}(R-r_{\rm min}(M))}{(R/r_s^{\rm MW})(1+R/r_s^{\rm MW})^2} \msun^{-1} \; {\rm kpc}^{-3},
\end{equation}
where a Heaviside function $\mathcal{H}(R-r_{\rm min}(M))$ accounts
for the tidal destruction of subhalos below a radius $r_{\rm min}(M)$.
The subhalo concentration-mass relation $P(c;M,R)$ is based on the
\citet{Bullock2001} model ($F=0.01$ and $K=3.75$) to which we have added a
radial dependence of the form
\begin{equation}
c_0^{\rm sub}(M,R) = c_0^{\rm B01}(M) \left( \f{R}{\rtwo} \right)^{-0.286} \hspace{-0.35in},
\end{equation}
which we have determined directly from our numerical simulations
\citep{Diemand2005b,Diemand2008}. With this scaling, subhalos at
$R_\odot$ are three times as concentrated as field halos. Note that we
include a log-normal scatter of width $\sigma_{\log_{10}c}=0.14$, such
that
\begin{equation}
P(c;M,R) = \f{1}{c \ln{10}} \; \f{1}{\sqrt{2\pi} \sigma_{\log_{10}c}} \; \exp{\left[ - \f{(\log_{10} c - \log_{10} c_0^{\rm sub})^2}{2 \; \sigma_{\log_{10}c}^2} \right]}.
\end{equation}

The flux in a solid angle $\Delta\Omega$ around $\psi$ originating
from a subhalo of mass M at a position $(\lambda,\theta,\phi)$ is
given by
\begin{equation}
  \mathcal{F}_{\rm halo}(M,c,\lambda,\theta,\phi,\Delta\Omega) = \int_{\Delta\Omega} d\!\cos\theta'\;d\phi' \int_{\rm l.o.s.} \hspace{-0.15in} \lambda'^2 \; \f{\rho^2(M,c,r(\lambda,\theta,\phi,\lambda',\theta',\phi'))}{4\pi \lambda'^2} \; d\lambda' ,
\label{eq:phi_halo}
\end{equation}
where the subhalo's density profile is given by an NFW profile with a
constant core below $r_c=10^{-8}$ kpc:
\begin{equation}
\rho(M,c,r) = \left\{
\begin{array}{ll}
\f{\rho_s}{(r/r_s)(1 + r/r_s)^2} & {\rm for} \, r > r_c \\
\f{\rho_s}{(r_c/r_s)(1 + r_c/r_s)^2} & {\rm for} \, r \leq r_c.
\end{array}
\right.
\end{equation}
The NFW scale density $\rho_s$ depends only on $c$, and the scale
radius $r_s$ on $M$ and $c$. We have checked that our results are not
strongly dependent on the value of $r_c$: $\mathcal{F}$ changes by less than
1\% for $r_c$ between $10^{-9}$ kpc and $10^{-7}$ kpc. 

Fig.\ref{fig:diagram} shows a diagram of the geometry relevant for the
calculation of $\mathcal{F}_{\rm halo}$. Without loss of generality we can
choose a coordinate system in which $\phi=0$ for the position
$\vec{p}$ of a given subhalo, and hence $\mathcal{F}_{\rm halo}$ is
independent of $\phi$. The distance $r$ from the center of the subhalo
at $\vec{p}=(\lambda, \theta, 0)$ to the point
$\vec{q}=(\lambda',\theta',\phi')$ within the cone of integration is
\begin{equation}
r(\lambda,\theta,\lambda',\theta',\phi') = \sqrt{\lambda^2 + \lambda'^2 -  2 \lambda \lambda' \left(\sin\theta \sin\theta' \cos\phi' + \cos\theta\cos\theta' \right)}.
\end{equation}
We are integrating over a very small opening angle $\Delta\Omega=4.36
\times 10^{-6}$ sr, corresponding to $\theta_{\rm max}=1.2\times
10^{-3}$, and hence we are justified in setting $\sin\theta=0$ and
$\cos\theta=1$, i.e. placing the subhalo on the line of sight given by
$\psi$. In this case $\mathcal{F}_{\rm halo}$ depends only on $M, c,
\lambda,$ and $\Delta\Omega$, and is given by
\begin{equation}
\mathcal{F}_{\rm halo}(M,c,\lambda,\Delta\Omega) = \f{\rho_s^2 r_s}{2} \int_{\cos\theta_{\rm max}}^1 \hspace{-0.25in} d\!\cos\theta' \int_0^\infty \!\! \f{1}{(\xi^2+\xi'^2-2\xi\xi'\cos\theta')\left(1+\sqrt{\xi^2+\xi'^2-2\xi\xi'\cos\theta'}\right)^4} \; d\xi',
\label{eq:Flux_halo}
\end{equation}
where $\xi=\lambda/r_s$ and $\xi'=\lambda'/r_s$.

\section{Calculation of detectable and extended subhalo mass functions}\label{sec:AppendixB}

The mass function of extended subhalos is given by
\begin{equation}
\f{dN_{\rm res}}{dM} = \int_{4\pi} \!d\cos\theta \; d\phi \int d\lambda \; \lambda^2 \int dc \; P(c;M,R) \; \f{dn_{\rm sub}}{dM}(M,R) \; \mathcal{H}(D_{\rm max}(M,c)-\lambda),
\end{equation}
where $\lambda$, $R$, $P(c;M,r)$, and $dn_{\rm sub}/dM$ are defined as in
Eq.~\ref{eq:phi1} (with $\psi=0$), $\mathcal{H}$ is the Heaviside step
function, and 
\begin{equation}
D_{\rm max} = r_s/9\;{\rm arcmin} = 10\;{\rm kpc} \f{1}{c(M,R)} \left( \f{M}{1\msun} \right)^{1/3}
\end{equation}
is the maximum distance out to which a subhalo of mass $M$ and
concentration $c(M,R)$ would appear extended. An additional constraint
on the surface brightness $\Phi$ can be incorporated by introducing a
second Heaviside step function, $\mathcal{H}(\Phi - \Phi_0)$.

\end{appendix}

\end{document}